\documentclass[PAPER, atlasdraft=false, texlive=2023, USenglish, cernpreprint, texmf, orcidlogo]{atlasdoc}
\usepackage{atlaspackage}
\usepackage{verbatim}
\usepackage{atlasbiblatex}

\usepackage[hepparticle, unit=false, hepprocess]{atlasphysics}

\usepackage{arydshln}

\graphicspath{{logos/}{figures/}}

\usepackage{ANA-TOPQ-2018-07-PAPER-defs} %
\AtlasTitle{
Measurement of single top-quark production in association with a $W$ boson
in $pp$ collisions at $\sqrt{s} = 13\,\text{TeV}$ with the ATLAS detector
}

\AtlasAbstract{
The inclusive cross-section for the production of a single top quark in association with a $W$ boson is measured using 140\,fb$^{-1}$ of proton--proton collision data collected with the ATLAS detector at $\sqrt{s} = 13$~TeV.
Events containing two charged leptons and at least one jet identified as originating from a $b$-quark are selected.
A multivariate discriminant is constructed to separate the $tW$ signal from the $t\bar{t}$ background.
The cross-section is extracted using a profile likelihood fit to the signal and control regions and it is measured to be $\sigma_{tW} = 75^{+15}_{-14}\,\text{pb}$, in good agreement with the Standard Model prediction.
The measured cross-section is used to extract a value for the left-handed form factor at the $Wtb$ vertex times the CKM matrix element $|f_\text{LV}V_{tb}|$ of $0.97 \pm 0.10$.
}

\author{The ATLAS Collaboration}

\AtlasRefCode{TOPQ-2018-07} %

\PreprintIdNumber{CERN-EP-2024-168}

\AtlasJournalRef{Phys. Rev. D 110 (2024) 072010}
\AtlasDOI{10.1103/PhysRevD.110.072010}

\hypersetup{pdftitle={ATLAS document},pdfauthor={The ATLAS Collaboration}}

\begin{document}

\maketitle

\tableofcontents

\clearpage

%
%
%

%

%
\section{Introduction}%
\begin{figure}[bp]
\centering
\includegraphics[width=0.45\linewidth]{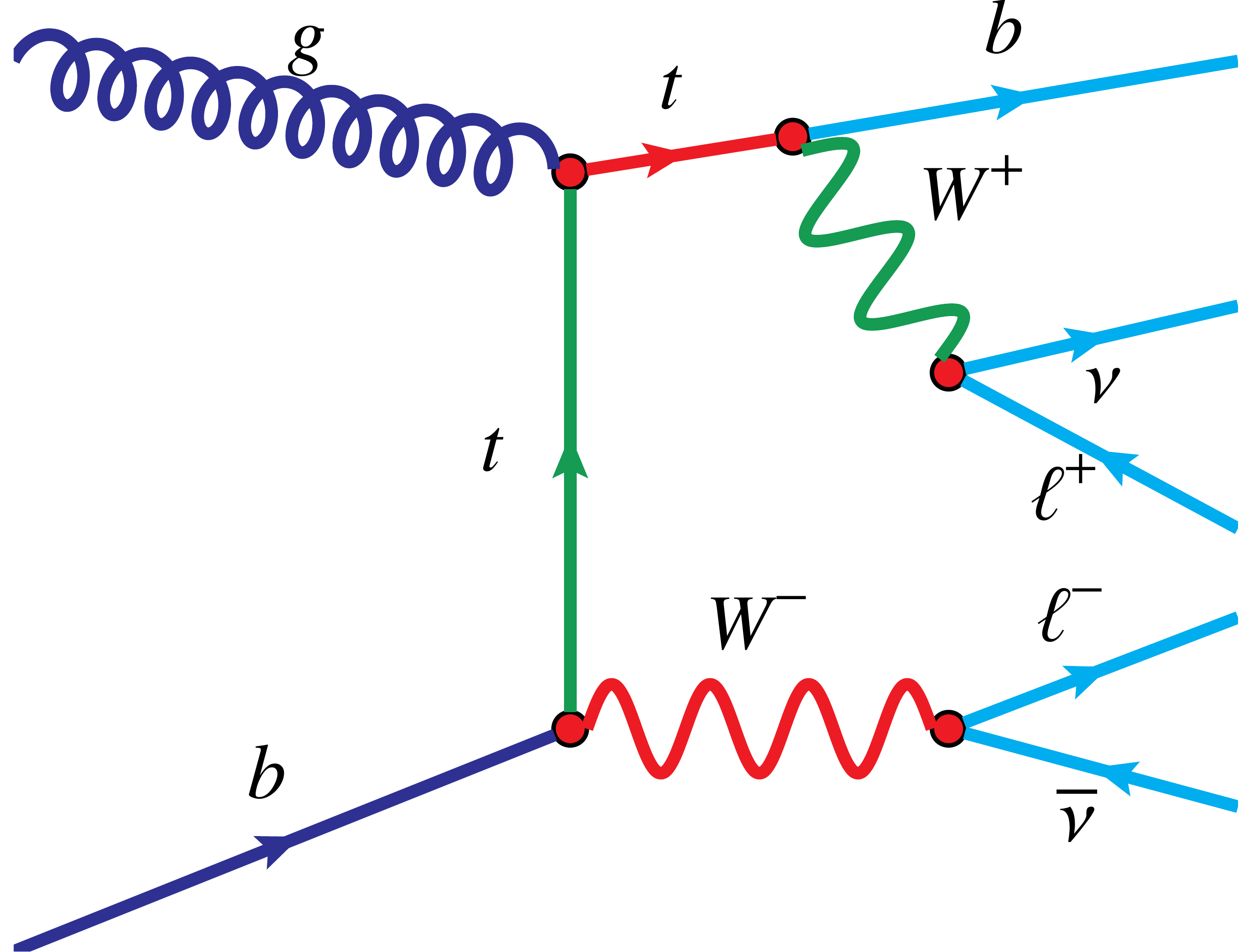}
\caption{
A representative leading-order Feynman diagram for the production of a single top quark in the \tW-channel
and then the decay of the $W$ boson and the top quark giving two leptons in the final state.
}
\label{fig:tW_prod}
\end{figure}

\label{sec:intro}
Top quarks are predominantly produced at the Large Hadron Collider (LHC)~\cite{Evans:2008zzb} in pairs via the strong interaction.
However, they can also be produced singly through processes that involve the electroweak $Wtb$ vertex, thus providing important tests of the electroweak interaction involving the third-generation quarks.
In the Standard Model (SM), single top-quark production at leading-order (LO) can be classified as proceeding through three distinct channels: $qb \rightarrow tq^\prime$ and $q\bar{q}^\prime \rightarrow t \bar{b}$ (and their charge conjugates) via $t$-channel and $s$-channel $W$-boson propagators, respectively, and $tW$ production, as shown in \cref{fig:tW_prod}.

At the LHC, the \tW-channel has the second largest
single top-quark production cross-section following
the $qb \rightarrow tq^\prime$ $t$-channel.
It represents approximately \SI{26}{\%} of the total single top-quark production
rate at \SI{13}{\TeV}, which makes it experimentally accessible at the LHC\@.
In addition, \tW production is the only single top-quark production mechanism that can be measured in the dilepton final state
with relatively clean signatures where the
only significant background is from top-quark pair production (\ttbar).

The cross-section for each of the three single top-quark production channels is primarily dominated by the coupling between the $W$ boson and the top and bottom quarks, which is proportional to the square of the Cabibbo-Kobayashi-Maskawa (CKM) matrix element $|V_{tb}|^{2}$~\cite{coupling_prop1,coupling_prop2}.
Single top-quark production therefore presents an opportunity to precisely test the structure of the SM, and to search for signs of physics beyond the SM (BSM) that can affect the $Wtb$ vertex.
The value of $|f_\text{LV}V_{tb}|^2$, where $f_\text{LV}$ is the left-handed form factor, was measured to be $1.015 \pm 0.031$ in the $qb \rightarrow tq^\prime$ $t$-channel~\cite{TOPQ-2019-02}.
In contrast to the $t$-channel and $q\bar{q}^\prime \rightarrow t \bar{b}$ $s$-channel processes, where BSM effects can arise from both the existence of four-fermion operators and corrections to the $Wtb$ vertex, the \tW-channel solely relies on the latter.
It is therefore crucial to investigate this channel independently to make a detailed comparison with the other channels~\cite{motiv1,motiv2}.
Furthermore, a precise measurement and good modeling of the \tW process is needed for many BSM searches~\cite{HDBS-2018-10,HIGG-2018-09} and  in Higgs boson and other top-quark
measurements, where it is one of the main background processes.

The predicted cross-section of $tW$ production in proton--proton ($pp$) collisions at a center-of-mass energy of 13~TeV is \(\sigma_{\text{\tW}}^{\text{theory}} = \SI[parse-numbers=false]{79.3 \numpmerr{+1.9}{-1.8}\,(\text{scale}) \pm 2.2\,(\text{PDF})}{\pb}\) and is computed at next-to-leading-order (NLO) in quantum chromodynamics (QCD) with the addition of third-order corrections from soft-gluon emissions by resumming next-to-next-to-leading-logarithmic (NNLL) terms~\cite{Kidonakis:2021vob}.
The top-quark mass is set to \qty{172.5}{\GeV} and the PDF4LHC21 set of parton distribution functions (PDF)~\cite{PDF4LHCWorkingGroup:2022cjn} is used.
The quoted uncertainty includes the uncertainty due to the choice of the renormalization scale $\mu_{r}$ and the factorization scale $\mu_{f}$, as well as the uncertainty in the PDFs.
The uncertainty in the scale choice is determined by varying the scales simultaneously up and down by a factor of two relative to the nominal value.
The PDF uncertainties are based on the Hessian method and include the uncertainty in the strong coupling constant \alphas.

Accurate estimates of rates and kinematic distributions of the \tW process are made more difficult at higher orders in \alphas due to
quantum interference with the \ttbar process.  Two commonly used approaches to separate \tW and \ttbar
production beyond LO are diagram removal (DR) and diagram subtraction (DS)~\cite{Frixione:2008yi}.  In the DR approach, all
doubly resonant NLO diagrams are removed from the calculation of the \tW
amplitude.
Instead, in the DS approach a subtraction term is implemented in the matrix-element calculation to locally cancel out the \ttbar doubly resonant
contributions. This interference effect was studied using \SI{13}{\TeV} collision data by the ATLAS Collaboration~\cite{TOPQ-2017-05}
and the results provide an important constraint on interference models.

At the LHC, evidence for the \tW process with \SI{7}{\TeV} collision data was presented by the ATLAS Collaboration~\cite{TOPQ-2011-17}
(with a significance of $\qty{3.6}{\sigma}$),
and by the CMS Collaboration~\cite{CMS-TOP-11-022} (with a significance of $\qty{4.0}{\sigma}$).
With \SI{8}{\TeV} collision data, CMS reported the observation of the \tW-channel with a significance of $\qty{6.1}{\sigma}$~\cite{CMS-TOP-12-040} while ATLAS observed it with a significance of $\qty{7.7}{\sigma}$~\cite{TOPQ-2012-20,TOPQ-2016-06}.
The \tW cross-section was also measured with \SI{13}{\TeV} collision data inclusively and differentially by the ATLAS~\cite{TOPQ-2015-16,TOPQ-2016-12} and CMS collaborations~\cite{CMS-TOP-17-018,CMS-TOP-20-002,CMS-TOP-21-010}.
In this paper, a measurement of \tW production is reported, extending previous measurements, by using a data sample collected with the ATLAS detector from 2015 to 2018 and corresponding to an integrated luminosity of
\SI{140}{\per\fb} of $pp$ collisions at a center-of-mass energy of $\sqrt{s} = \SI{13}{TeV}$.
After a brief introduction of the ATLAS detector in \cref{sec:detector}, further details of the data sample and the simulated samples used for the measurement are given in \cref{sec:datamc}.
The measurement is made using events containing at least one $b$-jet (identified as
containing a $b$-hadron) and exactly one electron and one muon with opposite charge in the final state, produced either directly from the decay of a \PW boson or from the decay of an intermediate $\tau$-lepton.
The \tW signal appears in this final state when the top quark decays into a \PW boson and a $b$-quark, with both the \PW bosons  decaying leptonically, as shown in \cref{fig:tW_prod}.
The $ee$ and $\mu\mu$ dilepton channels are not considered due to an additional large background from Drell--Yan $Z/\gamma^\ast \rightarrow e^+ e^-$ (or $\mu^+ \mu^-$) processes.
The object and event selection details are provided in \cref{sec:selection}.

Selected events are grouped into three different regions based on jet and $b$-jet multiplicities.
A boosted decision tree (BDT) analysis is performed in each of the three regions to build discriminants to separate the \tW signal from the dominant \ttbar background.
Details of the input variables and the BDT training are given in \cref{sec:bdt}, while the systematic uncertainties considered are discussed in \cref{sec:systematics}.
A simultaneous fit to the three regions is used to extract the signal strength, defined as the measured $tW$ cross-section over the SM prediction, and the \ttbar normalization.
The systematic uncertainties are included as nuisance parameters in the fit to data.
To avoid over-constraining some systematic uncertainties in non-physical ways, a limited range of the BDT discriminant is used in a profile-likelihood fit.
This approach enhances the reliability of the measurement, at the cost of a slightly larger uncertainty.
The fit procedure and the results are detailed in \cref{sec:fit} and \cref{sec:result} respectively,
and conclusions are given in \cref{sec:conclusion}.


%
%
%
%

%

%
\section{ATLAS detector}%
\label{sec:detector}
\newcommand{\AtlasCoordFootnote}{%
ATLAS uses a right-handed coordinate system with its origin at the nominal interaction point (IP)
in the center of the detector and the \(z\)-axis along the beam pipe.
The \(x\)-axis points from the IP to the centre of the LHC ring,
and the \(y\)-axis points upwards.
Polar coordinates \((r,\phi)\) are used in the transverse plane,
\(\phi\) being the azimuthal angle around the \(z\)-axis.
The pseudorapidity is defined in terms of the polar angle \(\theta\) as \(\eta = -\ln \tan(\theta/2)\) and is equal to the rapidity
$ y = \frac{1}{2} \ln \left( \frac{E + p_z c}{E - p_z c} \right) $ in the relativistic limit.
Angular distance is measured in units of \(\Delta R \equiv \sqrt{(\Delta y)^{2} + (\Delta\phi)^{2}}\).}

The ATLAS detector~\cite{PERF-2007-01} at the LHC covers nearly the entire solid angle around the collision point.\footnote{\AtlasCoordFootnote}
It consists of an inner tracking detector surrounded by a thin superconducting solenoid, electromagnetic and hadronic calorimeters,
and a muon spectrometer incorporating three large superconducting air-core toroidal magnets.

The inner-detector system (ID) is immersed in a \qty{2}{\tesla} axial magnetic field
and provides charged-particle tracking in the range \(|\eta| < 2.5\).
The high-granularity silicon pixel detector covers the vertex region and typically provides four measurements per track,
the first hit generally being in the insertable B-layer (IBL) installed before Run~2~\cite{ATLAS-TDR-19,PIX-2018-001}.
It is followed by the SemiConductor Tracker (SCT), which usually provides eight measurements per track.
These silicon detectors are complemented by the transition radiation tracker (TRT),
which enables radially extended track reconstruction up to \(|\eta| = 2.0\).
The TRT also provides electron identification information
based on the fraction of hits (typically 30 in total) above a higher energy-deposit threshold corresponding to transition radiation.

The calorimeter system covers the pseudorapidity range \(|\eta| < 4.9\).
Within the region \(|\eta|< 3.2\), electromagnetic calorimetry is provided by barrel and
endcap high-granularity lead/liquid-argon (LAr) calorimeters,
with an additional thin LAr presampler covering \(|\eta| < 1.8\)
to correct for energy loss in material upstream of the calorimeters.
Hadronic calorimetry is provided by the steel/scintillator-tile calorimeter,
segmented into three barrel structures within \(|\eta| < 1.7\), and two copper/LAr hadronic endcap calorimeters.
The solid angle coverage is completed with forward copper/LAr and tungsten/LAr calorimeter modules
optimized for electromagnetic and hadronic energy measurements respectively.

The muon spectrometer (MS) comprises separate trigger and
high-precision tracking chambers measuring the deflection of muons in a magnetic field generated by the superconducting air-core toroidal magnets.
The field integral of the toroids ranges between \num{2.0} and \qty{6.0}{\tesla\metre}
across most of the detector.
Three layers of precision chambers, each consisting of layers of monitored drift tubes, cover the region \(|\eta| < 2.7\),
complemented by cathode-strip chambers in the forward region, where the background is highest.
The muon trigger system covers the range \(|\eta| < 2.4\) with resistive-plate chambers in the barrel, and thin-gap chambers in the endcap regions.

The luminosity is measured mainly by the LUCID--2~\cite{LUCID2} detector that records Cherenkov light produced in the quartz windows of photomultipliers located close to the beampipe.

Events are selected by the first-level trigger system implemented in custom hardware,
followed by selections made by algorithms implemented in software in the high-level trigger~\cite{TRIG-2016-01}.
The first-level trigger accepts events from the \qty{40}{\MHz} bunch crossings at a rate below \qty{100}{\kHz},
which the high-level trigger further reduces in order to record complete events to disk at about \qty{1}{\kHz}.

A software suite~\cite{SOFT-2022-02} is used in data simulation, in the reconstruction
and analysis of real and simulated data, in detector operations, and in the trigger and data acquisition
systems of the experiment.


%
%
%
%

%

%
\section{Data and simulation}%
\label{sec:datamc}
Data were collected with the ATLAS detector from 2015 to 2018 in $pp$ collisions at a
center-of-mass energy of $\sqrt{s} = \SI{13}{\TeV}$ delivered by the LHC.
The full data sample corresponds to an integrated luminosity of $\SI{140}{\ifb}$.
Monte Carlo (MC) simulation is used to model the signal and background processes.
The effect of multiple interactions in the same and neighboring bunch crossings (pileup) is modeled
by overlaying the original hard-scattering event with simulated minimum-bias events generated with \PYTHIA[8.186]~\cite{Sjostrand:2007gs}
using the \NNPDF[2.3LO] PDF set~\cite{Ball:2012cx} and the A3 set of tuned parameters (tune)~\cite{ATL-PHYS-PUB-2016-017}.
The MC simulated events are weighted to reproduce the distribution of the interactions per bunch crossing observed in the data, referred to as \enquote{pileup reweighting}.
In all top-quark samples, the top-quark mass is set to $\mtop = \SI{172.5}{\GeV}$.
The decays of bottom and charm hadrons are simulated using the \EVTGEN program~\cite{EvtGen} for all non-\SHERPA samples.
Single-top $tW$  production is modeled using the \POWHEGBOX[v2]~\cite{Re:2010bp,Nason:2004rx,Frixione:2007vw,Alioli:2010xd}
generator which provides matrix elements at NLO in the strong coupling constant \alphas\
in the five-flavor scheme with the \NNPDF[3.0NLO] PDF set.
The functional form of the renormalization and factorization scale is set to the default scale, which is equal to the top-quark mass.
The diagram-removal scheme is used to treat the interference with \ttbar\ production~\cite{ATL-PHYS-PUB-2016-020}.
The events are interfaced with \PYTHIA[8.230]~\cite{Sjostrand:2014zea} using the A14 tune~\cite{ATL-PHYS-PUB-2014-021} and the \NNPDF[2.3LO] PDF set.
The inclusive cross-section is normalized to the theory prediction calculated at NLO in QCD with NNLL soft-gluon corrections~\cite{Kidonakis:2021vob}.
The \ttbar\ events are simulated using the \POWHEGBOX[v2]~\cite{Frixione:2007nw,Nason:2004rx,Frixione:2007vw,Alioli:2010xd} generator at NLO with the \NNPDF[3.0NLO] PDF and the \hdamp\ parameter\footnote{The \hdamp\ parameter controls the transverse momentum \pt\ of the first additional emission beyond the leading-order Feynman diagram in the parton shower and therefore regulates the  high-\pt\ emission against which the
\ttbar\ system recoils.} set to 1.5~\mtop~\cite{ATL-PHYS-PUB-2016-020}.
The events are interfaced with \PYTHIA[8.230] using the A14 tune~\cite{ATL-PHYS-PUB-2014-021} and the \NNPDF[2.3LO] PDF set.
The NLO \ttbar\ inclusive production cross-section is corrected to the theory prediction at next-to-next-to-leading-order (NNLO) in QCD including the resummation of next-to-next-to-leading-logarithmic (NNLL) soft-gluon terms calculated using \textsc{Top++2.0}~\cite{Beneke:2011mq,Cacciari:2011hy,Baernreuther:2012ws,Czakon:2012zr,Czakon:2012pz,Czakon:2013goa,Czakon:2011xx}.
The \ttbar\ samples are normalized to a reference cross-section of \qty[parse-numbers=false]{832 \numpmerr{+20}{-29} \pm 35}{\pb}, where the first uncertainty corresponds to QCD scale uncertainties and the second to PDF uncertainties.

Various samples are simulated to evaluate the modeling uncertainties in the  \tW and \ttbar production.
Further variations are obtained from the baseline sample, by using event weights to change the QCD factorization and renormalization scales, and the amounts of initial and final state radiation.
Alternative samples to evaluate the parton shower and hadronization uncertainties of \tW and \ttbar are produced with the \POWHEGBOX[v2] generator at NLO in QCD in the five-flavor scheme using the \NNPDF[3.0NLO] PDF\@.
The events are interfaced with \HERWIG[7.04]~\cite{Bahr:2008pv,Bellm:2015jjp}, using the H7UE tune~\cite{Bellm:2015jjp} and the \MMHT[LO] PDF set~\cite{Harland-Lang:2014zoa}.

To assess the uncertainty in the hard scattering simulation and the matching scheme, additional \tW and \ttbar samples are generated with the \MGNLO generator at NLO in QCD in the five-flavor scheme, interfaced with \PYTHIA[8.230].

Previous studies have seen improved agreement between data and prediction in \ttbar events, particularly for the top-quark \pT distribution, when comparing with NNLO calculations~\cite{TOPQ-2018-17}.
Therefore, all \ttbar samples are reweighted to match their top-quark \pT distribution to that predicted at NNLO in QCD and NLO electroweak (EW) accuracy~\cite{Czakon:2017wor}.

To estimate the uncertainty in the interference between \tW and \ttbar\ production,
an alternative \tW \POWPY~8 sample is simulated using the diagram-subtraction scheme.

Backgrounds with minor contributions, referred to as minor backgrounds in the following, are also simulated.
The production of a vector boson in association with jets, $V$+jets ($V=W,Z$) is simulated with the \SHERPA[v2.2.1]~\cite{Bothmann:2019yzt} generator.
In this setup, NLO-accurate
matrix elements for up to two jets, and LO-accurate matrix elements for up
to four jets are calculated with the Comix~\cite{Gleisberg:2008fv} and
\OPENLOOPS~\cite{Cascioli:2011va,Denner:2016kdg} libraries. They are matched
with the \SHERPA parton shower~\cite{Schumann:2007mg} using the MEPS@NLO
prescription~\cite{Hoeche:2011fd,Hoeche:2012yf,Catani:2001cc,Hoeche:2009rj}.
Diboson samples, with fully leptonic and semileptonic decays, are simulated with the
\SHERPA[v2.2.2]~\cite{Bothmann:2019yzt} generator. In this setup multiple
matrix elements are matched and merged with the \SHERPA parton shower
based on Catani--Seymour dipole factorization~\cite{Gleisberg:2008fv,Schumann:2007mg}
using the MEPS@NLO prescription.
The virtual QCD corrections for matrix elements at NLO accuracy are
provided by the \OPENLOOPS library.
These \SHERPA samples are simulated using the
\NNPDF[3.0NNLO] set, along with the dedicated set of
tuned parton-shower parameters developed by the \SHERPA authors.

The MC simulated samples used for the nominal estimates were processed through a simulation~\cite{SOFT-2010-01} of the ATLAS detector based on \GEANT~\cite{Agostinelli:2002hh}, while those used for evaluating systematic uncertainties were processed with a fast simulation that relies on a parameterization of the calorimeter response~\cite{ATL-PHYS-PUB-2010-013}.


%
%
%
%

%

%
\section{Object and event selection}%
\label{sec:selection}

This section describes the reconstruction of the basic objects used in the analysis and how events are selected using these objects.
Events must first be selected by a single-electron trigger~\cite{TRIG-2018-05} or a single-muon trigger~\cite{TRIG-2018-01}.
The lowest \pt threshold used for electrons was \SI{24}{\GeV} (\SI{26}{\GeV}) in 2015 (2016--2018), while for muons the threshold was \SI{20}{\GeV} (\SI{26}{\GeV}).

\subsection{Object reconstruction}

The primary vertex is selected as the $pp$ vertex candidate that has at least two tracks and the highest sum of the squared transverse momenta of all the matched tracks with $\pT > \SI{500}{MeV}$~\cite{ATL-PHYS-PUB-2015-026}.

Electron candidates are reconstructed from energy deposits in the electromagnetic (EM) calorimeter matched with ID tracks~\cite{EGAM-2018-01}. The
clusters are required to be in the pseudorapidity range $|\eta| < 2.47$, with the transition region between the barrel and endcap EM
calorimeters, $1.37 < |\eta| < 1.52$, excluded. The candidate electrons are required to have transverse energy
$\ET > \SI{20}{\GeV}$. Further requirements on the electromagnetic shower shape, calorimeter energy to tracker momentum ratio, and other
discriminating variables are combined into a likelihood-based object quality cut, to enhance electron selection efficiency while rejecting
backgrounds from photon conversions and hadrons misidentified as electrons~\cite{EGAM-2018-01}.
Furthermore, electrons are required to be isolated from other activity in the event using criteria based on
the calorimeter energy in a cone of radius $\Delta R= 0.2$ around the electron, %
and the sum of the \pt of tracks in a cone with size varying with electron \pt. %
The efficiency of the isolation requirement depends on the electron \pt.
For example, in $\PZ\rightarrow\Pe\Pe$ decays, it is $90\%$ ($99\%$) efficient for electrons with $\pt > \SI{25}{\GeV}$ ($\pt > \SI{60}{\GeV}$).
Electron tracks are also required to be consistent with coming from the primary vertex requiring $|d_0/\sigma_{d_{0}} |<5$ and $|z_0 \sin \theta|< \SI{0.5}{mm}$.
Here $d_0$ and $\sigma(d_0)$ are the transverse impact parameter and its uncertainty, and $z_0$ is the longitudinal impact parameter relative to the primary vertex of the event along the beam line.
Muon candidates are identified by matching reconstructed ID tracks with tracks in the muon spectrometer~\cite{MUON-2018-03},
and are required to have $|\eta| < 2.5$ and $\pT > \SI{20}{\GeV}$.
To reduce background from heavy flavor decays into muons inside jets, muons are required to satisfy analogous isolation requirements to electrons.
Muon tracks are also required to be consistent with the primary vertex requiring $|d_0/\sigma_{d_{0}} |<3$ and $|\Delta z_0\sin \theta|< \SI{0.5}{mm}$.
Similarly to electrons, correction factors are applied to simulated muons to account for small differences between data and simulation in the identification efficiencies.

Jets are reconstructed using a particle-flow algorithm~\cite{PERF-2015-09}, using calorimeter measurements for the energies of neutral particles
and ID track-momentum measurements for charged hadrons.
Jet reconstruction starts with particle-flow objects
and are formed using the \antikt algorithm
with radius parameter of $R = 0.4$~\cite{Cacciari:2008gp,Fastjet}.
These jets are then calibrated to the particle level by the application of a jet energy scale derived from simulation.
The jet energy is further corrected by applying {\em in situ} corrections for the contribution from pileup events based on $\sqrt{s} = \SI{13}{\TeV}$ data~\cite{JETM-2018-05}.
Selected jets are required to have $\pt > \SI{20}{\GeV}$ and $|\eta| < 2.5$ to ensure all of the matched charged particles are inside the coverage of the ID\@.
To suppress jets arising from pileup, a discriminant called the jet-vertex-tagger (JVT)~\cite{PERF-2014-03} is constructed using a two-dimensional likelihood method.
The JVT discriminates between jets produced in the hard-scatter and pileup processes, and is
required to be larger than $0.59$ for the jets with $\pT < \SI{60}{\GeV}$ and $|\eta|<2.4$, corresponding to \SI{92}{\%} efficiency in \ttbar events.

Jets containing $b$-hadrons ($b$-jets) are identified using the \texttt{DL1r}~\cite{FTAG-2019-07} algorithm, which uses a deep feed-forward neural network to calculate the probability that a jet is a $b$-jet, $c$-jet, or light jet (jets containing neither a $b$- or $c$-hadron).
The inputs to this $b$-tagging algorithm include impact parameter and reconstructed secondary vertex information from tracks in the jet.
Candidate $b$-jets are required to have a $b$-tagging discriminant value greater than a threshold that corresponds to
a $b$-tagging efficiency of \SI{77}{\%} in simulated \ttbar events.
With this requirement the rejection factors are 170 for light jets and five for $c$-jets.
Correction factors are applied to account for remaining differences between data and simulation
separately for $b$-jets, $c$-jets and light jets.
The missing transverse momentum, with magnitude \met, serves as an indicator of the transverse momentum attributed to undetected neutrinos, computed as the magnitude of the negative vector sum of the momenta of all identified particles (electrons, photons, muons, \Pgt-leptons, and jets) in the event transverse plane~\cite{JETM-2020-03}.
It also includes a contribution from soft hadronic activity by using
reconstructed charged-particle tracks matched with the primary vertex but not with any of the reconstructed objects.

To avoid cases where the detector response to a single physical object is
reconstructed as two separate final-state objects, an overlap-removal procedure is employed.
First, if an electron candidate and a muon candidate share a track, the electron candidate is removed.
Next, the closest jet to each electron within a $\Delta R$ distance of 0.2
is removed to reduce the portion of electrons being reconstructed as jets.
Electrons within distances of $0.2 \le \Delta R < 0.4$ from any of the remaining non-pileup
jets are removed to reduce backgrounds from non-prompt, non-isolated electrons resulting from heavy flavor hadron decays.
Jets with fewer than three matched tracks and a distance $\Delta R < 0.2$ from a muon candidate are then
removed to reduce fake jets from muons depositing energy in the calorimeters.
Finally muons that are a distance of $0.2 \le \Delta R < 0.4$ from any
of the surviving jets are removed to avoid contamination with non-prompt muons from heavy flavor hadron decays.

\subsection{Event selection and categorization}

After the objects are identified, a selection is applied to maximize the purity of \tW signal events while reducing major background contributions from \ttbar and minor contributions from $\PZ+\text{jets}$, $\PW+\text{jets}$, and diboson production.
Simulated events with at least one selected electron or muon which is misidentified or non-prompt (i.e., \ matched to a $b$- or $c$-hadron decay, a misidentified jet, a kaon or pion decay, or an electron from a photon conversion) are separated out from all available MC samples and are merged to estimate the misidentified/non-prompt background.
The contribution where both the leptons are misidentified/non-prompt was determined to be negligible.
The misidentified/non-prompt background estimate is validated using \Pepm{}\Pgmpm same-charge regions with the same jet multiplicities, where a larger contribution of such background is present.

The \tW final state used for this measurement comprises two oppositely charged leptons ($e^\pm \mu^\mp$ events) and missing transverse momentum from the two $W$ boson decays, and one $b$-jet from the top-quark decay. %
The event preselection begins by requiring an oppositely charged electron--muon pair (dilepton \Pepm{}\Pgmmp events).
The leading lepton in the event must satisfy $\pT > \SI{27}{\GeV}$, while an event is rejected if a third lepton is
present with $\pT > \SI{20}{\GeV}$. One of the selected leptons must be matched to the trigger object used to select the event.
At least one jet with $\pt>\SI{25}{\GeV}$, $\abseta<2.5$, and $b$-tagged
at the \SI{77}{\percent} efficiency criterion discussed above, is required in each event.

After preselection, the dilepton events are categorized based on the jet and \Pqb -jet multiplicities.
At LO the signal process results in a final state with one $b$-jet arising from the top-quark decay,
while the \ttbar{} process results in two $b$-jets from the two top-quark decays.
Events with one additional jet are also considered due to the expected contributions to the \tW signal
mostly from QCD radiation in higher-order processes.

Based on the expected final states, three orthogonal event categories are defined:
events with exactly one selected jet that is also \Pqb-tagged (denoted \texttt{1j1b}),
events with exactly two selected jets one of which is \Pqb-tagged (\texttt{2j1b}),
and events with exactly two jets where each are $b$-tagged (\texttt{2j2b}).
The third category helps to constrain the \ttbar{} background normalization.
These three categories, \texttt{1j1b}, \texttt{2j1b}, and \texttt{2j2b}, are referred to as fit regions,
as they are used in the simultaneous fit procedure described in \Sect{\ref{sec:fit}}.

The expected event yields after preselection, separated according to fit region, for the signal
and background processes with their total uncertainties before fit, are shown in \Tab{\ref{tab:presel_yields}}.
Also shown is the number of observed data events in each fit region.
Agreement between data and simulation in all regions is observed.
\begin{table}[htbp]
\sisetup{group-minimum-digits=3}
\centering
\caption{
Expected and observed number of events in each fit region of the analysis after preselection.
The \tW and \ttbar processes are normalized to their respective theoretical cross-sections.
The uncertainties include statistical and systematic uncertainties while the uncertainties stemming from the theoretical calculation of the \tW and \ttbar processes are not included.
}
\label{tab:presel_yields}
\begin{tabular}{l
S[table-format=6.0]@{$\,\pm\,$}S[table-format=5.1]
S[table-format=6.0]@{$\,\pm\,$}S[table-format=5.1]
S[table-format=6.1]@{$\,\pm\,$}S[table-format=5.1]
}
\toprule
& \multicolumn{2}{c}{\texttt{1j1b}} &  \multicolumn{2}{c}{\texttt{2j1b}}  & \multicolumn{2}{c}{\texttt{2j2b}}  \\
\midrule
$tW$               &  27500 &  2400  &  17500 & 1900   & 5400   & 900 \\
$t\bar{t}$         & 111000 & 11000  & 182000 & 12000  & 155000 & 14000 \\
$Z+$jets           &   1810 &   230  & 1000   & 120    & 87     & 19 \\
Diboson            &    820 &   180  & 840    & 190    & 23.2   & 3.2 \\
Misidentified/non-prompt lepton    &    430 &   220  & 800    & 400    & 180    & 90 \\
\midrule
Total prediction   & 141000 & 12000 & 202000  & 12000  & 160000 & 14000 \\
\midrule
Data               & \multicolumn{2}{c}{\num{139349}} & \multicolumn{2}{c}{\num{199095}} & \multicolumn{2}{c}{\num{158314}} \\
\bottomrule
\end{tabular}
\end{table}


\FloatBarrier

%
%
%

%

%
\section{Separation of signal from background}%
\label{sec:bdt}
After the event preselection described in \Sect{\ref{sec:selection}}, the selected dilepton events
consist primarily of \ttbar and \tW with minor contributions from other processes as listed in \Tab{\ref{tab:presel_yields}}.

To better separate events from \tW{} production and those from \ttbar, Boosted Decision Trees (BDTs) are used, based on the implementation provided by the LightGBM software package~\cite{bdt:lgbm}.
One BDT is trained for each analysis fit region, with simulated \tW{} events serving as signal and simulated \ttbar{}
events serving as background. Other minor backgrounds are not included in the training procedure.

Reconstructed final-state objects (jets, leptons, \met) are used to construct various kinematic quantities.
The definitions of these variables are given in \Tab{\ref{tab:variables}}. For a given object type, the subscript $1$ is used to denote the object with highest $\pt$, referred
to as the leading object. For example, $\ptsys (\ell_1\ell_2j_1\MET)$ is the $\pt$ of the vector sum of momenta of the leading lepton, subleading lepton, leading jet, and $\MET$.
Although jets are required to have $\pt > \SI{25}{\GeV}$ to define the fit regions, the transverse momentum of additional soft jets ($j_S$) in events is considered, where the soft jets are defined by having $\SI{20}{\GeV} < \pt  < \SI{25}{\GeV}$.
In cases where no soft jet is present, a value of zero is used.
\begin{table}[htbp]
\begin{center}
\caption{Definitions of variable names constructed from the set of final-state objects $s = (o_1, ..., o_n)$ for $n \ge 1$.
For variables that compare two sets of objects, $s_1$ and $s_2$, $\vec{s}_i$ is the momentum-vector sum of the objects in set $s_i$.
}
\label{tab:variables}
\begin{tabular}{ll}
\toprule
Variable & Definition \\
\midrule
$\ptsys (s)$                                         &  Transverse component of the vector sum of momenta \\
$m(s)$                                                & Invariant mass of the system of multiple objects $s$ \\
$H(s)$                                                & Scalar sum of momenta \\
$\HT(s)$                                             & Scalar sum of transverse momenta \\
Centrality$(s)$                                   &  Centrality of the system $s$, given by $\HT(s) / H(s)$  \\
$\Delta R (\vec{s}_1,\ \vec{s}_2)$      &  $\eta - \phi$ separation between $\vec{s}_1$ and $\vec{s}_2$  \\
$\Delta p_{\mathrm{T}} (\vec{s}_1,\ \vec{s}_2)$ &   Magnitude of the transverse component of $\vec{s}_1 - \vec{s}_2$ \\
$\HT^{\text{ratio}}(s_1,\ s_2)$         &  Ratio of the $\HT$ of the two systems: $\HT(s_1) / \HT(s_2)$ \\
$p_{\mathrm{T}}(j_{S1})$                   &  \pt of the leading soft jet \\
$m_{\mathrm{T}}(o_1 \MET)$            & Transverse mass of object $o_1$ and $\MET$: $m_{\mathrm{T}} = \sqrt{2\pt(o_1)\MET(1 - \cos \Delta \phi (o_1, \MET))}$    \\
\bottomrule
\end{tabular}
\end{center}
\end{table}

For each fit region, an optimization procedure maximizes the area under the receiver operating characteristic curve (AUC) value while reducing overtraining. The signal and
background samples are separated into three categories for the optimization procedure: training set (40\% of each sample), validation set (20\% of each sample), and testing set (40\% of each sample).
A grid search is performed to find the optimal hyperparameters of the BDT model.

Two strategies are used to prevent overtraining in parallel with the grid search optimization.
The first strategy is the practice of early stopping.
Early stopping is a mechanism to end the training procedure if the AUC value of the validation set stops improving.
The second strategy is to do a two-sample Kolmogorov-Smirnov (KS) test~\cite{bdt:KStestKolmogorov,bdt:KStestSmirnoff} on the training and test sets.
The KS test is performed on the signal and background
samples (comparing the training and testing distributions). If one of
the signal or background KS-test $p$-values is less than 0.05, the
specific grid point is rejected.
The final hyperparameter configuration is chosen by selecting the training configuration with the highest area under the curve after rejecting any overtrained models.
The final hyperparameter configuration for each region is reported in \Tab{\ref{tab:bdt_hyperparams}}.

\begin{table}[htbp]
\begin{center}
\caption{The final hyperparameter settings of the BDTs trained in
each analysis region.}
\label{tab:bdt_hyperparams}
\begin{tabular}{lcccc}
\toprule{Region} & Learning rate & Number of leaves & {Minimum data in a leaf} & {Maximum depth} \\
\midrule
\texttt{1j1b} & 0.2 & 20 & 50 & 4 \\
\texttt{2j1b} & 0.1 & 20 & 120 & 7 \\
\texttt{2j2b} & 0.2 & 20 & 50 & 4 \\
\bottomrule
\end{tabular}
\end{center}
\end{table}

\begin{figure}[htbp]
\centering
\subfloat[]{\includegraphics[height=0.17\textheight]{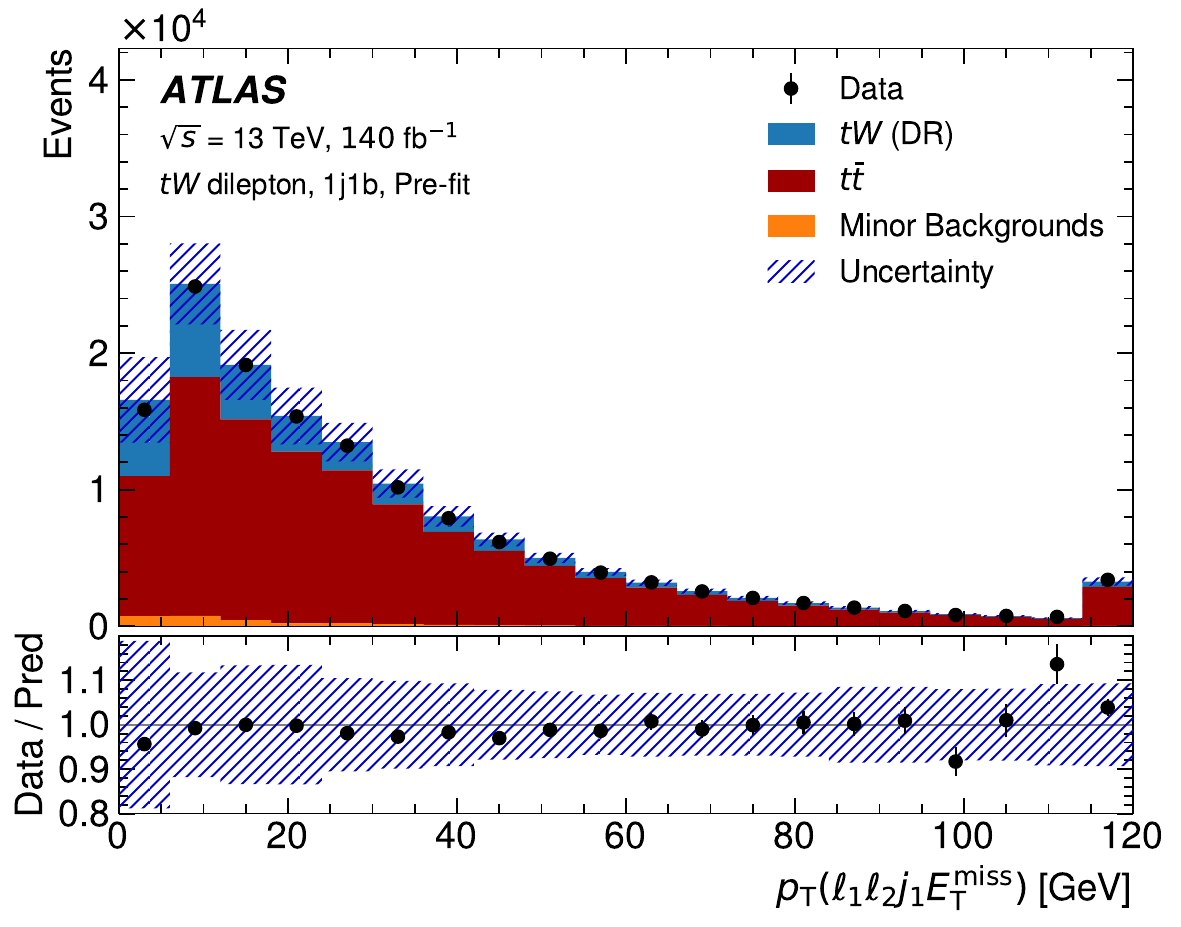}}
\subfloat[]{\includegraphics[height=0.17\textheight]{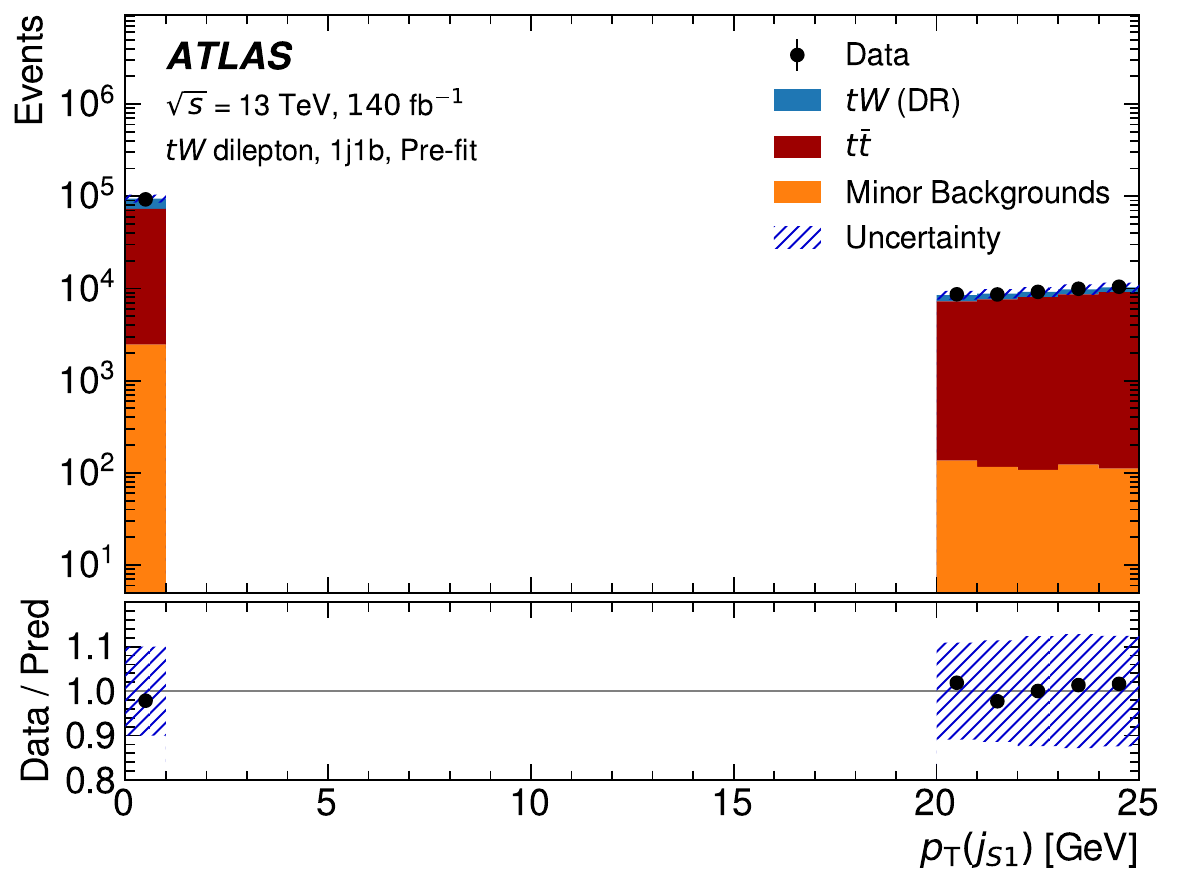}}
\subfloat[]{\includegraphics[height=0.17\textheight]{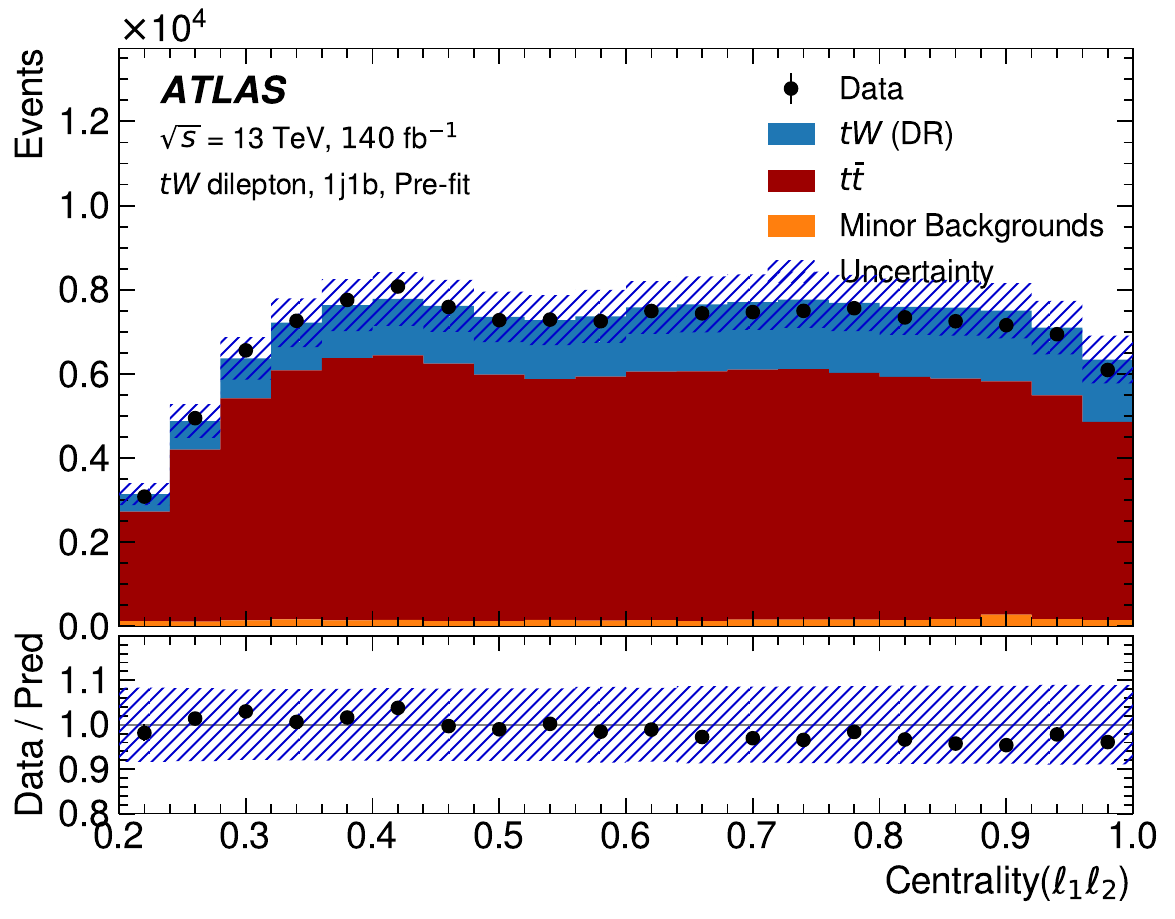}} \\
\subfloat[]{\includegraphics[height=0.17\textheight]{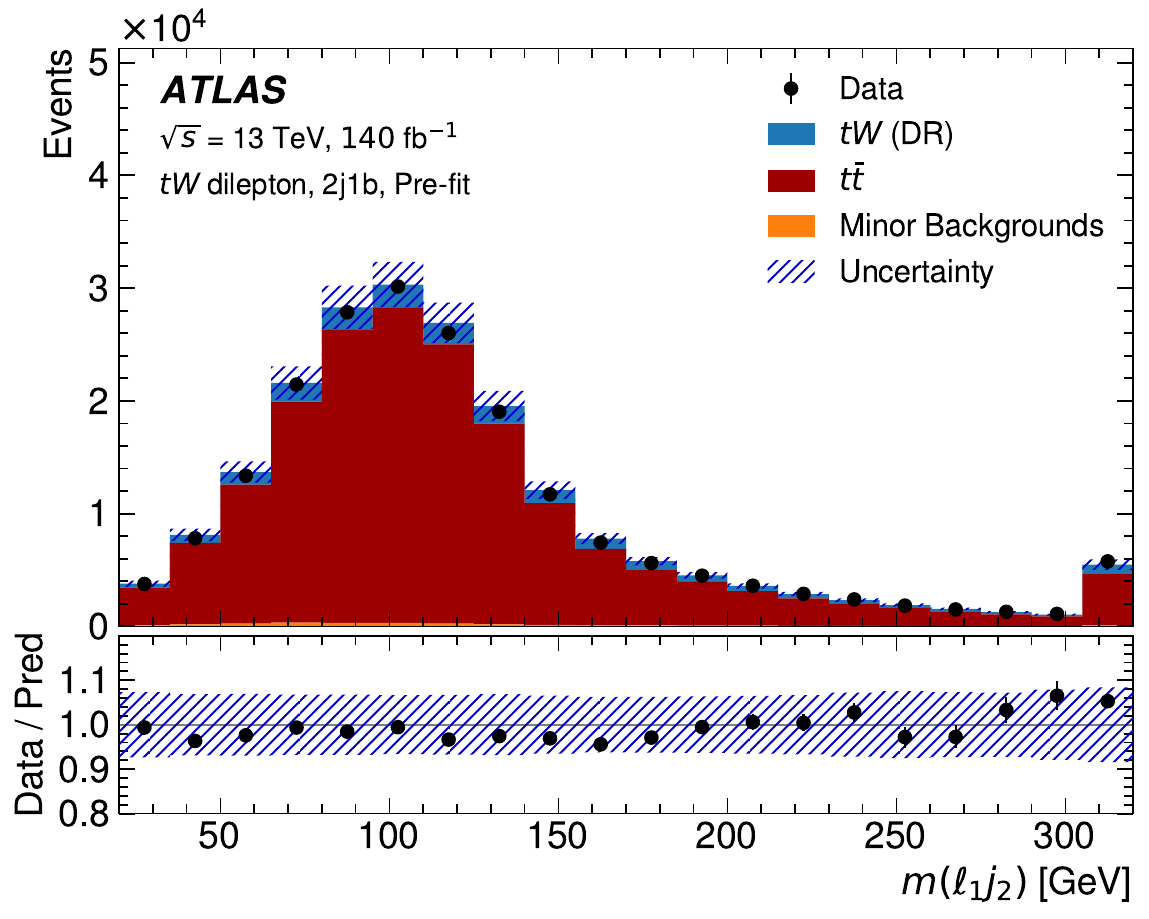}}
\subfloat[]{\includegraphics[height=0.17\textheight]{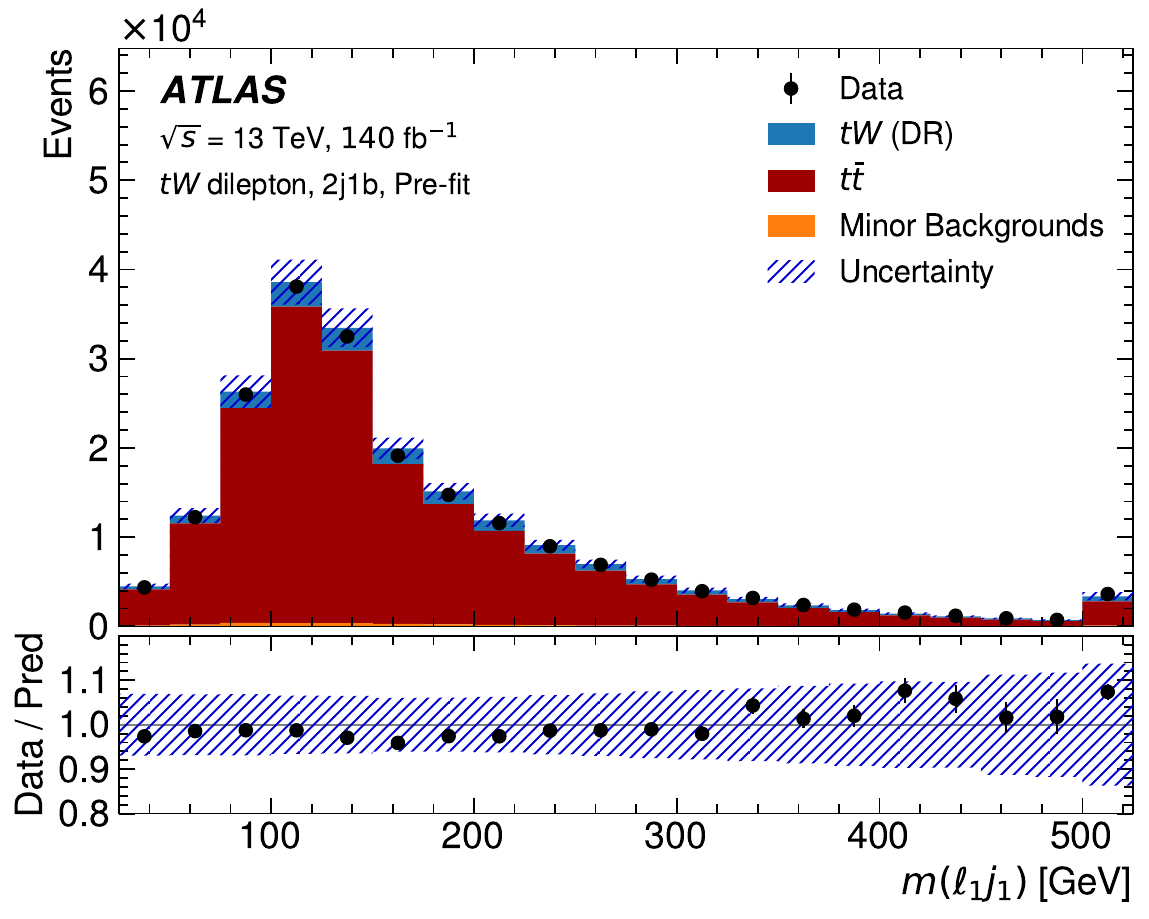}}
\subfloat[]{\includegraphics[height=0.17\textheight]{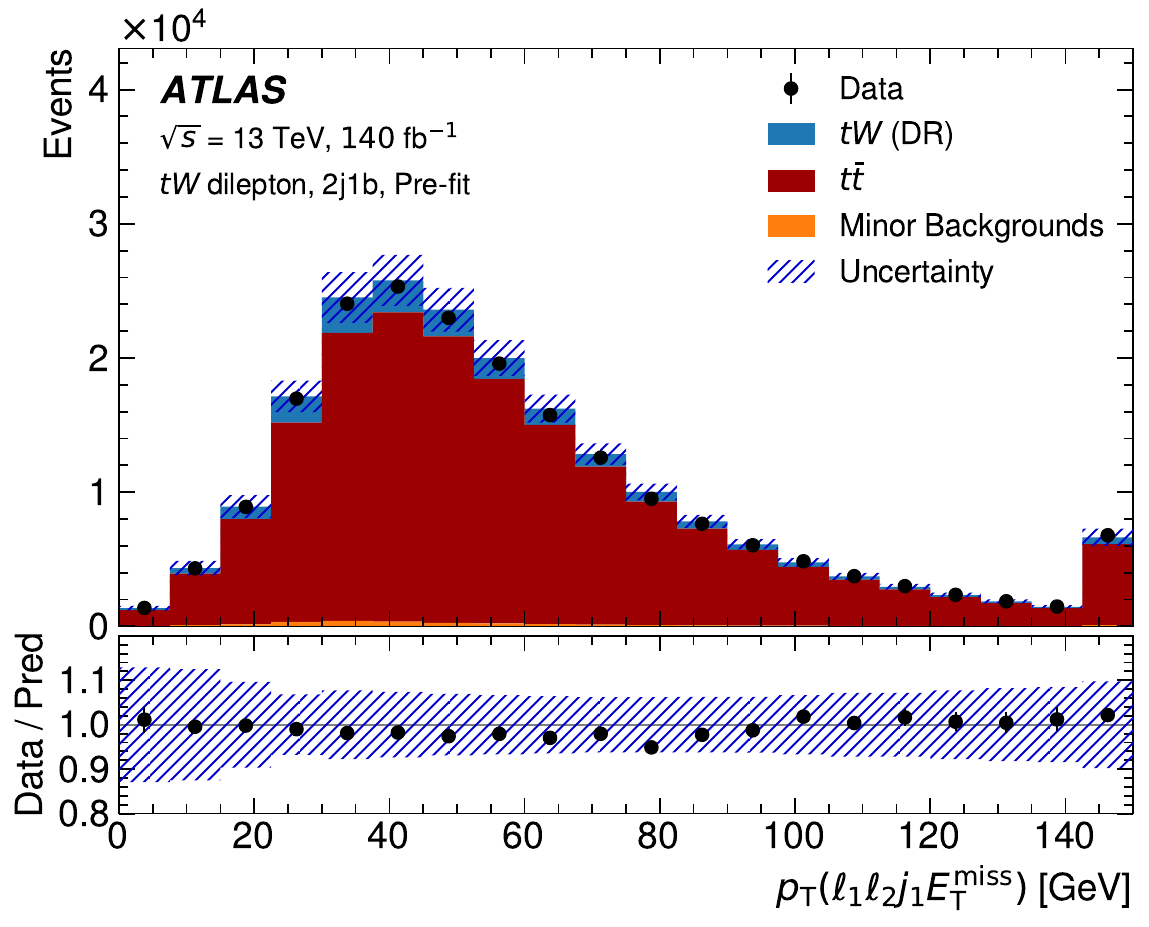}} \\
\subfloat[]{\includegraphics[height=0.17\textheight]{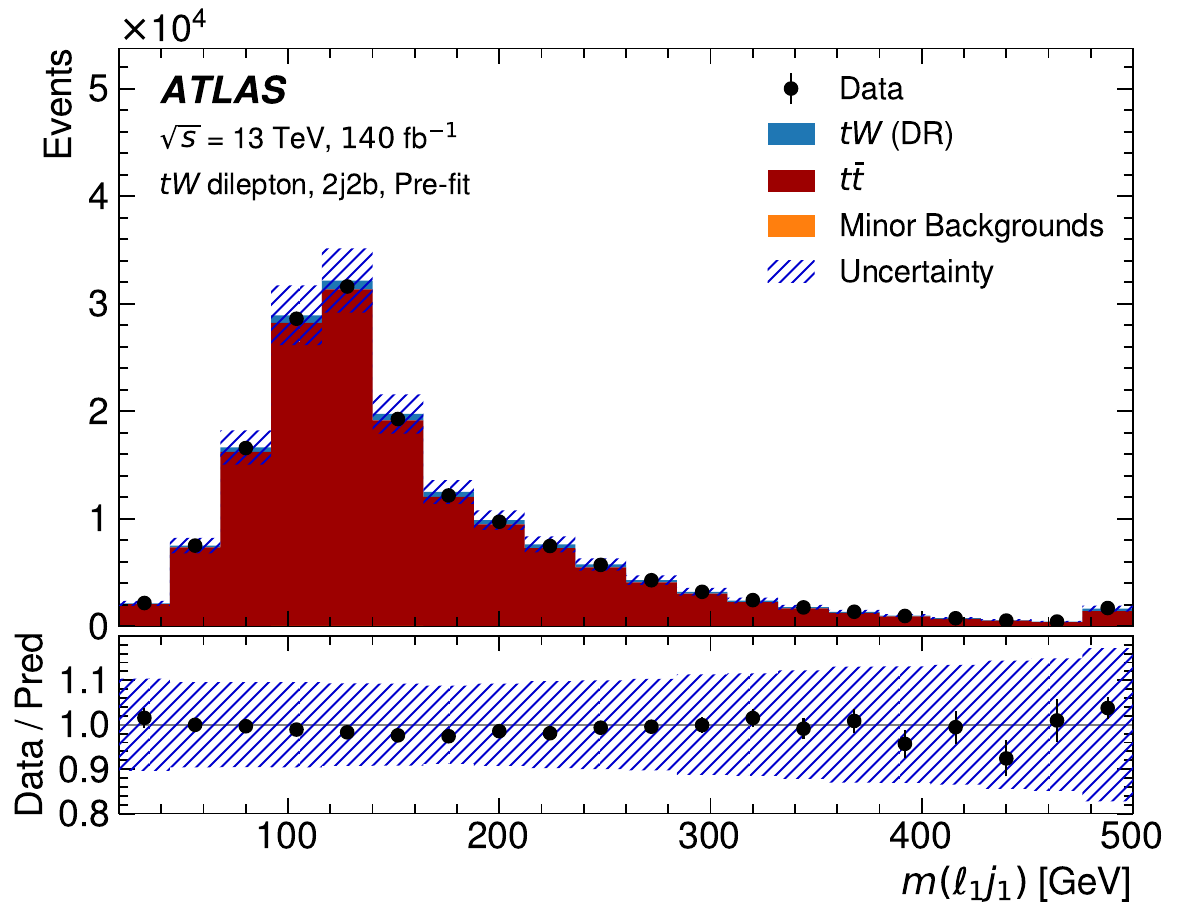}}
\subfloat[]{\includegraphics[height=0.17\textheight]{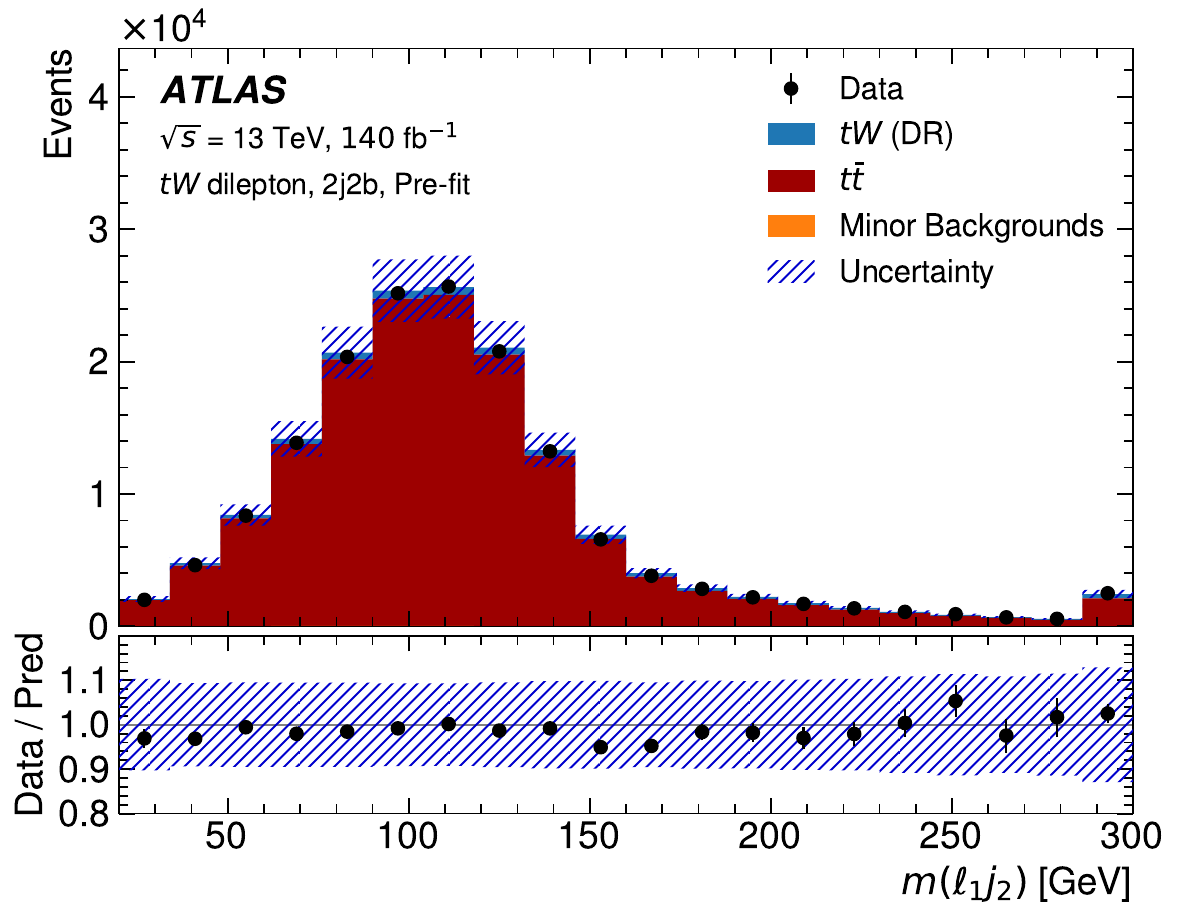}}
\subfloat[]{\includegraphics[height=0.17\textheight]{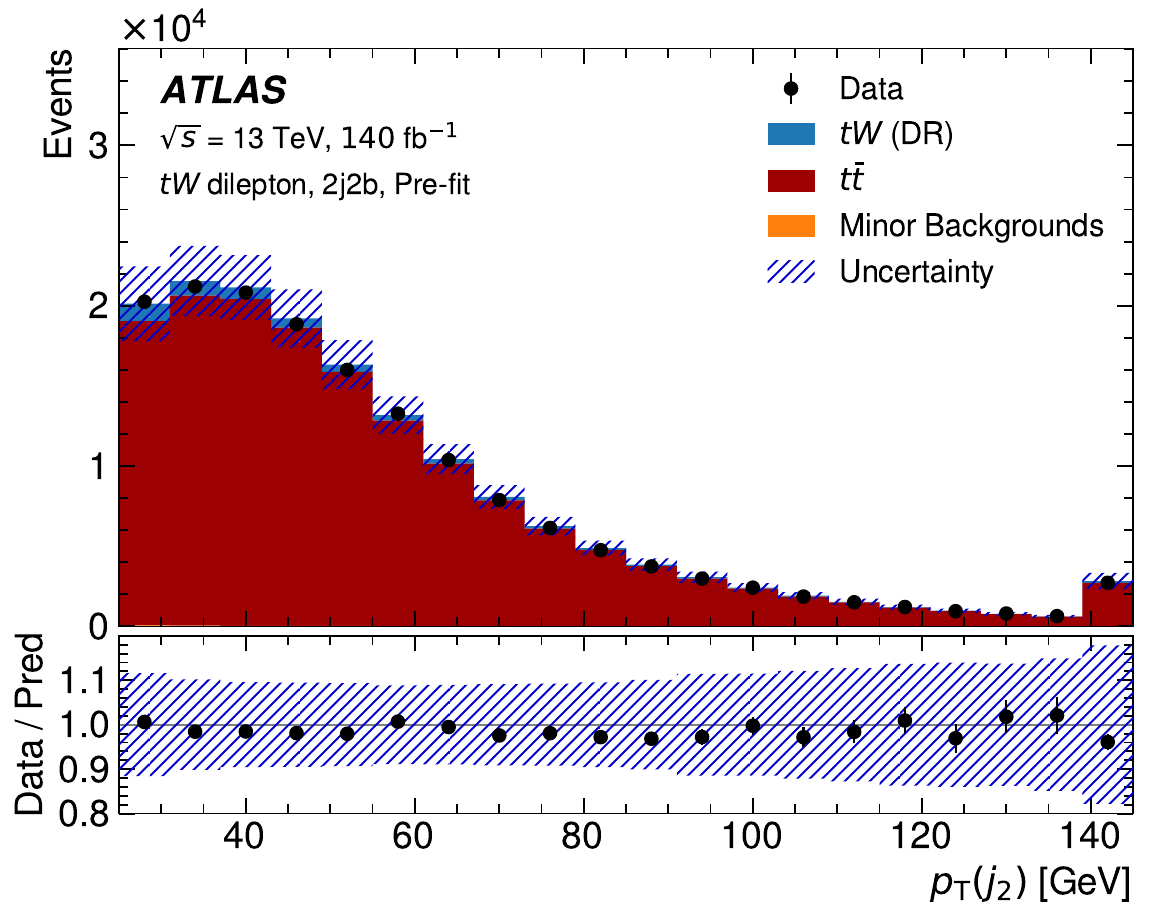}} \\
\caption{
BDT input distributions for the three highest ranked variables (defined in \cref{tab:bdt_imp_all_regions}) from the \texttt{1j1b} region ((a)--(c)), the \texttt{2j1b} region ((d)--(f)), and the \texttt{2j2b} region ((g)--(i)) before the fit.
\caprest{}
\capuncertaintybandprefit{}}
\label{fig:top3_inputs_prefit}
\end{figure}

\begin{figure}[htbp]
\centering
\includegraphics[width=1.0\linewidth]{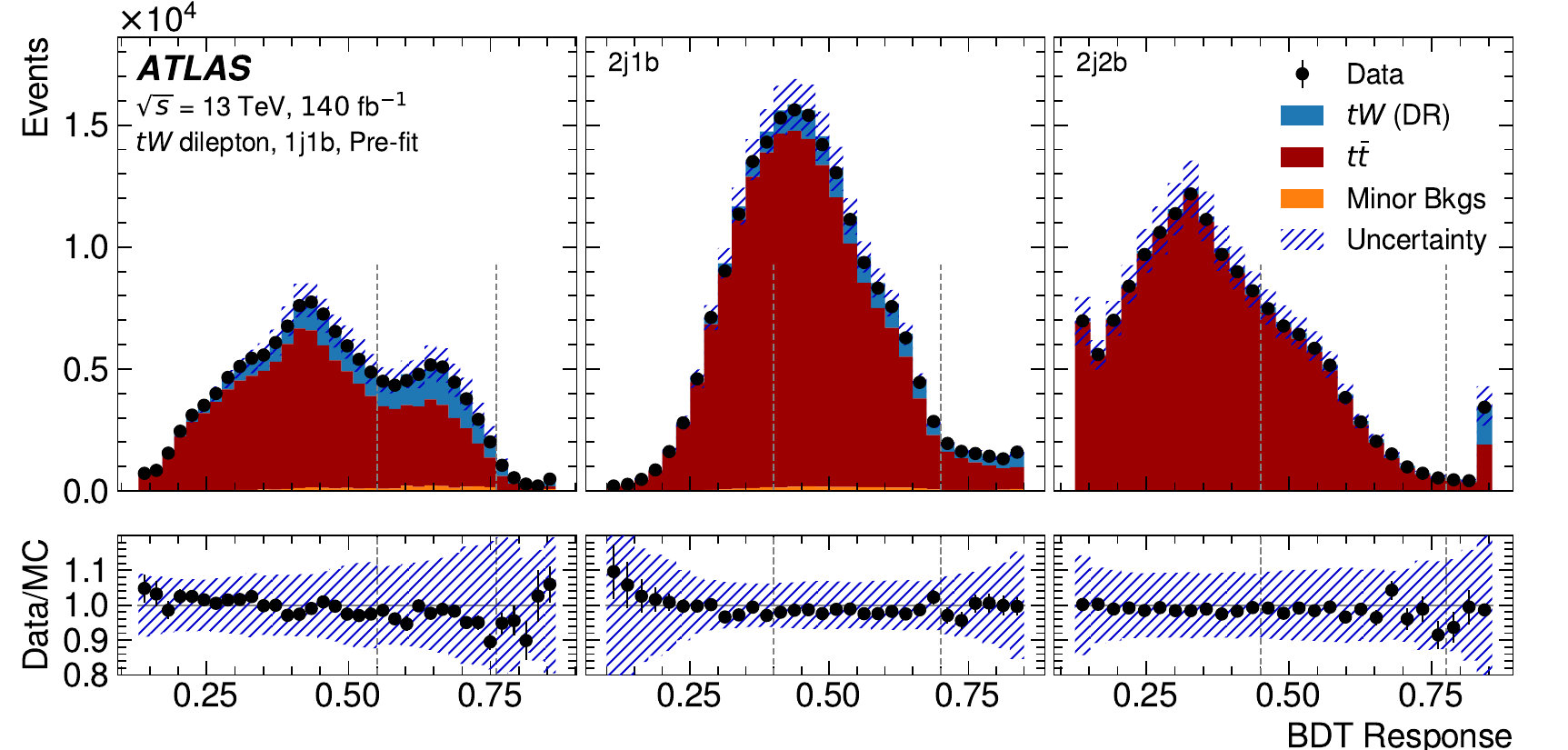}
\caption{
The pre-fit BDT response distributions for the three regions used in the fit.
The BDT range used in the fit described in \cref{sec:fit} is indicated by vertical dashed lines.
The first and last bins include the underflow and overflow, respectively.
\capuncertaintybandprefit}
\label{fig:prefit_dists}
\end{figure}

The list of variables used and their  rank for all fit regions is given in \cref{tab:bdt_imp_all_regions}.
The number of variables is significantly smaller and thus the training procedure is simplified; compared with using the complete set of variables, a performance loss of less than 2\% of the maximum performance is achieved.
LightGBM can report the importance based on how much the performance is improved by using a particular variable.
The top three variables from each region according to this importance metric is given where the sum of gains for all variables is normalized to one.
Distributions of the three highest ranked variables for each region are shown in \cref{fig:top3_inputs_prefit}.
The BDT outputs are shown in \cref{fig:prefit_dists}.
The pre-fit uncertainty shown in each distribution includes the complete set of uncertainties described in \Sect{\ref{sec:systematics}}.
The BDT range used in the fit described in \cref{sec:fit} is indicated by vertical dashed lines.
The MC predictions describe the data well, consistent with the total systematic uncertainties.

\begin{table}[htbp]
\begin{center}
\caption{
Importance ranks for the variables used in the BDT for all three regions.
The variable name definitions are given in \Tab{\ref{tab:variables}}.
The variables without a rank are not used in that region and \enquote{--} indicates undefined variables in that region.
}
\label{tab:bdt_imp_all_regions}
\begin{tabular}{lccc}
\toprule
Variable & \texttt{1j1b} & \texttt{2j1b} & \texttt{2j2b} \\
\midrule
$\ptsys (\ell_1\ell_2j_1\MET)$                   & 1  & 3 &  \\
$p_{\mathrm{T}}(j_{S1})$                         & 2  &    &   \\
Centrality($\ell_1\ell_2$)                       & 3  &    &    \\
$m_{\mathrm{T}}(j_1 \MET)$                       & 4               &    &    \\
$m(\ell_1j_1)$                                   & 5               &  2  &  1 \\
$m(\ell_2j_1)$                                   & 6  &     &  4 \\
$\Delta p_{\mathrm{T}} (\ell_1, \ell_2)$         & 7  &     &   \\
$\ptsys (\ell_1\ell_2)$                          & 8  &  5   &  \\
$m(\ell_2j_1\MET)$                               & 9  &     &  \\
$\ptsys (j_1\MET)$                               & 10 &    &  \\
$\Delta R (\ell_2, j_1)$                         &      &  6  & \\
$\ptsys (\ell_1\ell_2\MET)$                      &    &    & 6 \\
$m(\ell_1j_2)$                                   & --   & 1 &  2 \\
$\ptsys (\ell_1\ell_2j_1j_2\MET)$                & --   &  4 &  \\
$\HT^{\text{ratio}}(\ell_1\ell_2, \ell_1\ell_2j_1j_2\MET)$\ \ \ \  & -- &  7 & \\
$\HT^{\text{ratio}}(\ell_1\ell_2, \ell_1\ell_2j_1\MET)$ & -- &  8  &  \\
$p_{\mathrm{T}}(j_2)$                            & --  &  & 3 \\
$m(\ell_2j_2)$                                   & -- &   &  5 \\
\bottomrule
\end{tabular}
\end{center}
\end{table}

\FloatBarrier
%
%
%

%
\section{Systematic uncertainties}%
\label{sec:systematics}
Systematic uncertainties are divided into experimental and theoretical sources.
Each uncertainty is assigned a Gaussian distributed nuisance parameter that allows the uncertainty to be constrained by data.

\subsection{Experimental uncertainties}
\label{subsec:expsys}
Sources of experimental uncertainty include the uncertainties in the lepton efficiencies, the lepton energy scale and resolution, the \met soft-term calculation, the jet energy scale (JES) and resolution (JER), the JVT, the $b$-tagging efficiency, the pileup reweighting and the luminosity measurement.
Among these, the dominant sources of uncertainty are those in the determination of JES and JER.

The differences between the electron (muon) trigger, reconstruction and isolation efficiencies in data and those in MC simulation are corrected for by scale factors derived from dedicated $\PZ\to\Pe\Pe$ ($\PZ\to\Pgm\Pgm$) enriched control samples using a tag-and-probe method~\cite{EGAM-2018-01,MUON-2018-03}.

The JES is calibrated using simulation and in situ techniques~\cite{JETM-2018-05}. %
The JES uncertainty is decomposed into a set of 30 uncorrelated components.
The contributing effects are from pileup, jet flavor composition and response, single particle detector response, $\eta$ intercalibration, \textit{in situ} measurement, and the properties of jets not completely contained inside of the calorimeter volume.
The JER uncertainty is represented by eight components accounting for differences between data and MC simulation~\cite{JETM-2018-05}.
The uncertainties in the $b$-tagging calibration are determined separately for $b$-jets, $c$-jets, and light-flavor jets~\cite{FTAG-2018-01,FTAG-2020-08,FTAG-2019-02}.
A total of 19 components, accounting for differences between data and simulation, are used (nine for $b$-jets, four for $c$- and light-jets, and two for the MC-based extrapolation to high-\pt jets).
These uncertainty
factors are applied as changes to the scale factor weights on a
per-event basis.

The energy scale and resolution uncertainties for hard objects (leptons and jets) are propagated to the \met through
the re-computation of its corresponding terms. Therefore, the impact is evaluated when the event selections are re-applied after having
shifted the lepton or jet energy.
The soft track related uncertainties are derived from comparisons between data and MC simulation of the \pT balance between the hard and
soft \met components using $\PZ+$jets events~\cite{JETM-2020-03}.

The uncertainty in the combined 2015--2018 integrated luminosity is \SI{0.83}{\percent}~\cite{DAPR-2021-01}, derived from the calibration of the luminosity scale using $x$-$y$ beam-separation scans, and using the LUCID-2 detector for the baseline luminosity measurements~\cite{LUCID2}.

\subsection{Theoretical uncertainties}
Uncertainties stemming from theoretical models are estimated by comparing a set of predicted distributions produced with different modeling assumptions.

One of the most important theoretical uncertainties is the interference between \tW and \ttbar processes.
The uncertainty is evaluated by comparing the DR and DS schemes as implemented in \POWHEGBOX[2] while the \ttbar sample remains unchanged.
The uncertainty is symmetrized.

The uncertainty due to the choice of the matching scheme is assessed by comparing the nominal generator setup with an alternative setup for \ttbar and \tW, using \MGNLO~\cite{Alwall:2014hca} with the \NNPDF{3.0NLO} PDF set for the hard-scattering calculation. These are interfaced with \PYTHIA{8.230} using the A14 tune and the \NNPDF{23LO} PDF set. This uncertainty is symmetrized.

The uncertainty in parton showering and hadronization is evaluated by comparing \POWHEGBOX[2]+ \PYTHIA[8] and \POWHEGBOX[2]+ \HERWIG[7] samples for  \ttbar and \tW. %
A total normalization uncertainty is constructed from the difference between the total yields predicted by the two models.
A migration uncertainty is constructed by normalizing the overall yields of the two samples to be the same
and comparing the yields in individual regions (one nuisance parameter for \tW and
one for \ttbar).
Shape uncertainties are defined in each of the three analysis regions and assumed to be uncorrelated.

Initial and final state radiation (ISR/FSR) uncertainties are evaluated
using internal weights stored in the nominal \ttbar and \tW \POWHEGBOX[2]+\PYTHIA[8] samples.
The uncertainty due to missing higher-order QCD corrections in the matrix-element computation is estimated by independently varying the renormalization and factorization scales by factors of 2.0 and 0.5 relative to the central value.
The nominal tuning used in the parton shower generator (called the A14 tune) is also varied up (called the \textsc{Var3cUp} variation) and down (called the \textsc{Var3cDown} variation).
The FSR uncertainty is evaluated by varying the scale $\mu_\text{FSR}$
by factors of 2.0 and 0.5 relative to the central value.
These uncertainties are taken to be uncorrelated across regions and samples.
The uncertainty due to the scale choice for matching the matrix-element calculation of the \ttbar process to the parton shower is estimated by using an additional \ttbar sample produced with the \hdamp parameter set to 3.0~\mtop instead of 1.5~\mtop, while keeping all other generator settings the same as in  the nominal sample of \ttbar events.
This uncertainty is treated uncorrelated across regions and labeled \enquote{hdamp} in the following.

The PDF uncertainties are evaluated using the 30 eigenvectors of the PDF4LHC15 combined PDF set~\cite{Butterworth:2015oua}, taking into account the different effects on the \ttbar, \tW, $Z+$jets and inclusive $Z$ processes and their correlations.

For the \ttbar process, a one-sided symmetrized uncertainty is constructed in all analysis regions by comparing the nominal distribution after the top-quark~\pt reweighting with the distribution before the reweighting.

The modeling uncertainties in the minor background normalizations are estimated as follows.
An overall uncertainty of \SI{11}{\percent} is applied to $\PZ+$jets events in the two regions with one
$b$-jet, and a \SI{17}{\percent} uncertainty is applied in the region with two $b$-jets~\cite{STDM-2017-38}.
A \SI{22}{\percent} uncertainty is applied to diboson events with one $b$-jet, and a \SI{13}{\percent} uncertainty used for diboson events with
two $b$-jets, based on comparisons between \SHERPA and \POWHEG+\PYTHIA models~\cite{TOPQ-2015-09}.
A \SI{50}{\percent} uncertainty is applied to the normalization of the backgrounds from non-prompt and misidentified leptons~\cite{TOPQ-2015-09}.
All minor backgrounds normalization uncertainties are uncorrelated across regions, and found to have a relatively minor impact on the final measurements
due to the small contributions from these processes.


\FloatBarrier
%
%
%

%
\section{Fit strategy}%
\label{sec:fit}
BDT distributions of the observed data, simulated samples, and systematic uncertainty templates, are used together in a binned profile-likelihood fit to extract the measured \tW and \ttbar cross-sections.
The fit  to all of the bins of all three regions is used to  calculate the expected yields of each process.
The core components of the likelihood function are Poisson distribution terms that account for the probability of observing the expected yields in data, and additional (Gaussian distribution ) terms parameterized by nuisance parameters that account for variations from the central value of each source of systematic uncertainty.
The strength parameters for both the processes, $\mu_{tW}$ and $\mu_{t\bar{t}}$, defined as the ratio of the measured cross-section to that of the theoretical prediction, are free parameters of the fit.
The strength parameters for all other (minor) processes are Gaussian distributed and constrained by their respective systematic uncertainties to their SM expectations.

Sources of uncertainty are removed from the fit if their effects are determined to be below a certain threshold.
Specifically, a threshold of \SI{0.05}{\percent} is applied to normalization effects, and for shape effects, at least one bin must vary by more than \SI{0.1}{\percent}.

A challenge presented by the DR versus DS uncertainty arises from its nature as a two-point systematic uncertainty, making its constraint difficult to justify.
Given that the DS scheme includes doubly resonant diagrams while the DR scheme does not, the DS sample tends to contain more events similar to \ttbar.
Selectively excluding certain ranges of the BDT response distributions helps mitigate the stringent constraint while retaining a similar level of sensitivity, benefiting from the large available data sample.
Therefore, additional cuts on the BDT response are applied to reduce the large impact of this uncertainty and its undesired constraint by the fit.

The impact of another two-point systematic uncertainty, the \ttbar parton shower uncertainty, in this measurement is notable as \ttbar is the dominant background.
The showering algorithm can lead to variations in the kinematic properties and final-state particles of the \ttbar events and consequently on the BDT distribution.
Therefore, the BDT distribution range is further reduced by also taking the parton showering effects into account.

This optimization alleviates the stringent constraints imposed on the nuisance parameters of both the \tW DR vs DS uncertainty and the \ttbar parton shower uncertainty, changing the post-fit uncertainty to the pre-fit uncertainty ratios from an initial \SIrange[range-phrase={--}]{30}{40}{\%} across fit regions to a more relaxed \SIrange[range-phrase={--}]{60}{90}{\%}.
The post-fit error of $\mu_{tW}$ increases from \SI{13}{\%} to \SI{19}{\%} with these selections on the BDT responses.
The fractions of \tW and \ttbar events satisfying the additional selection cuts are \SI{53}{\%} and \SI{41}{\%}, respectively.
The final fit range for each region can be seen in \cref{fig:postfit_dists}.

\section{Results}%
\label{sec:result}
\subsection{Measurement of the inclusive cross-section}

The observed and predicted yields after the BDT cuts in the three regions are shown in
\cref{tab:prefit_yields}.
Good agreement is observed between data and prediction.
\begin{table}[htbp]
\centering
\caption{
Yields of each process, and total yields, before and after the fit, in each fit region.
The pre-fit uncertainties include statistical and systematic uncertainties while the uncertainties stemming from the theoretical calculation of the \tW and \ttbar processes are not included.
The post-fit uncertainties include statistical and systematic uncertainties,
as well as the uncertainties in $\mu_{tW}$ and $\mu_{\ttbar}$.
The post-fit errors in the total prediction are reduced compared with the sum in quadrature of the individual uncertainties due to correlations resulting from the fit.}
\label{tab:prefit_yields}
\begin{tabular}{l
S[table-format=5.0]@{$\,\pm\,$}S[table-format=4.1]
S[table-format=5.0]@{$\,\pm\,$}S[table-format=4.1]
S[table-format=5.1]@{$\,\pm\,$}S[table-format=4.1]
}
\toprule
& \multicolumn{2}{c}{\texttt{1j1b}}  & \multicolumn{2}{c}{\texttt{2j1b}}  & \multicolumn{2}{c}{\texttt{2j2b}}  \\
\midrule
Pre-fit $tW$         & 13000 & 1400 & 11900  & 1200 & 2000  & 400 \\
Pre-fit $t\bar{t}$   & 28000 & 4000 & 112000 & 8000 & 43000 & 4000 \\
Pre-fit $Z+$jets     & 1130  & 160  & 750    & 100  & 38    & 12 \\
Pre-fit diboson      & 380   & 80   & 570    & 130  & 8.5   & 1.3 \\
Pre-fit non-prompt   & 140   & 70   & 450    & 220  & 54    & 27 \\
\midrule
Pre-fit total prediction & 43000 & 5000 & 126000 & 8000 & 45000 & 4000 \\
\midrule
\midrule
Post-fit $tW$        & 12500 & 2000 & 11400  & 2200 & 2000  & 400 \\
Post-fit $t\bar{t}$  & 27400 & 2000 & 110300 & 2200 & 42100 & 500 \\
Post-fit $Z+$jets    & 1100  & 120  & 750    & 80   & 38    & 6 \\
Post-fit diboson     & 380   & 80   & 570    & 120  & 8.6   & 1.1 \\
Post-fit non-prompt  & 140   & 70   & 450    & 220  & 53    & 27 \\
\midrule
Post-fit total prediction & 41600 & 210 & 123500 & 400 & 44150 & 210 \\
\midrule
Data   & \multicolumn{2}{c}{\num{41591}} & \multicolumn{2}{c}{\num{123531}} & \multicolumn{2}{c}{\num{44149}} \\
\bottomrule
\end{tabular}
\end{table}

\begin{figure}[htbp]
\centering
\includegraphics[width=0.7\linewidth]{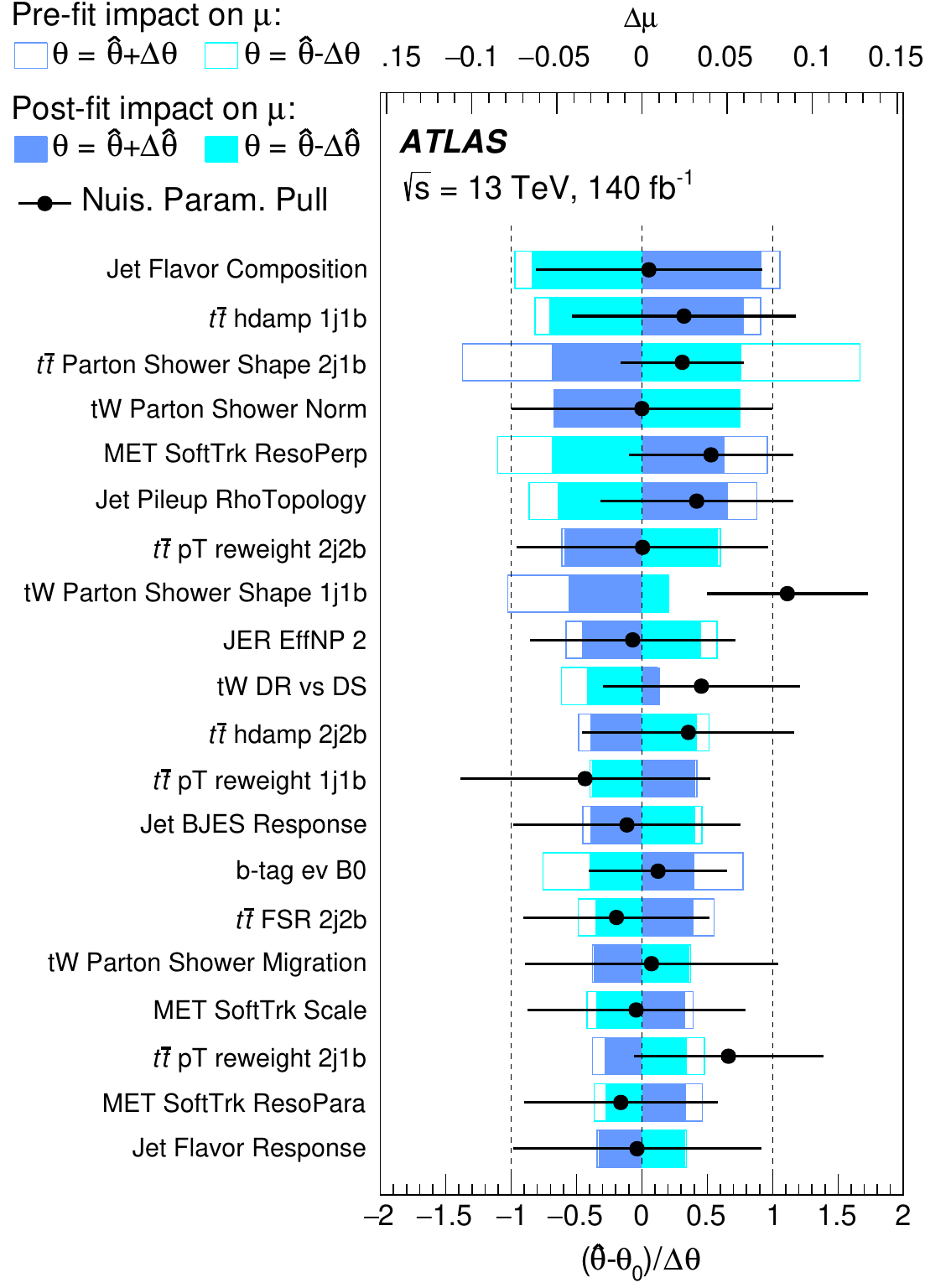}
\caption{
Ranking of the top 20 nuisance parameters by their most significant post-fit impact on $\mu_{\tW}$ in the fit,
after applying the additional BDT selection criteria shown in \cref{fig:prefit_dists}.
The empty and filled blue rectangles represent respectively the pre-fit and post-fit impacts on $\mu_{\tW}$ and are referring to the upper scale.
The impact of a nuisance parameter, $\Delta \mu$, is computed by comparing the nominal best-fit value of $\mu_{\tW}$ with the fit result obtained when fixing the respective nuisance parameter to its best-fit value, $\hat{\theta}$, shifted by its pre-fit (post-fit) uncertainties $\pm \Delta \theta$ ($\pm \Delta \hat{\theta}$).
The black points show the pulls of the nuisance parameters relative to their nominal values, $\theta_0$.
The pulls and their corresponding post-fit errors, expressed as $\Delta \hat{\theta} / \Delta \theta$, are also depicted on the lower scale.
\enquote{MET SoftTrk ResoPerp (ResoPara)} denotes the \met uncertainty in the soft-track \pt resolution smearing perpendicular (parallel) to the \met direction.
}
\label{fig:ranking_obs}
\end{figure}

The ten nuisance parameters with the largest post-fit impact on $\mu_{\tW}$ are shown in \cref{fig:ranking_obs}.
Due to the additional BDT selection criteria applied, the impact of the DR vs DS uncertainties are significantly reduced as is reflected in the ranking.
No unacceptable constraint is observed and the only significant pull comes from the \tW parton shower in the \texttt{1j1b} region.
The dominant uncertainties are in the  signal and background modeling, jet reconstruction, and \MET.
Some of these modeling uncertainties are estimated from a comparison between two alternative MC generators that crucially provide a conservative coverage of the uncertainty stemming from the modeling of the processes.

After fitting to data the observed signal strengths are $\mu_{\tW} = {\fittWDataValue} \numpmerr{\fittWDataErrorUp}{\fittWDataErrorDn}$ for \tW
and  $\mu_{\ttbar} = \fitttDataValue\numpmerr{\fitttDataErrorUp}{\fitttDataErrorDn}$ for \ttbar.
The fit value for $\mu_{\tW}$ corresponds to a measured cross-section of:
\begin{equation*}
\sigma_{\tW} = \fittWXSValue\numpmerr{\fittWXSErrorUp}{\fittWXSErrorDn}~\text{pb} = \fittWXSValue
\pm\fittWXSErrorStat\,\text{(stat.)}
\numpmerr{\fittWXSErrorSystUp}{\fittWXSErrorSystDn} \,\text{(syst.)}
\pm\fittWXSErrorLumi\,\text{(lumi.)}~\unit{\pb},
\end{equation*}
to be compared with the theory prediction of \(\sigma_{\text{\tW}}^{\text{theory}} = 79.3 \numpmerr{+1.9}{-1.8}\,(\text{scale}) \pm 2.2\,(\text{PDF})\,\unit{\pb}\).
The additional BDT selection criteria do not introduce any bias in the final result, which has an uncertainty of $\SI{19}{\%}$, compared to $\SI{13}{\%}$ without these criteria applied.
They make the final result less sensitive to modeling uncertainties and assumptions, resulting in a more robust measurement.
The observed (expected) significance is \significanceData\ (\significanceAsimov) standard deviations calculated using the asymptotic approximation~\cite{Cowan:2010js}.
The uncertainty in the measured cross-section is reduced by around \SI{40}{\%} compared with the previous measurement~\cite{TOPQ-2015-16}.

\begin{figure}[htbp]
\centering
\includegraphics[width=1.0\linewidth]{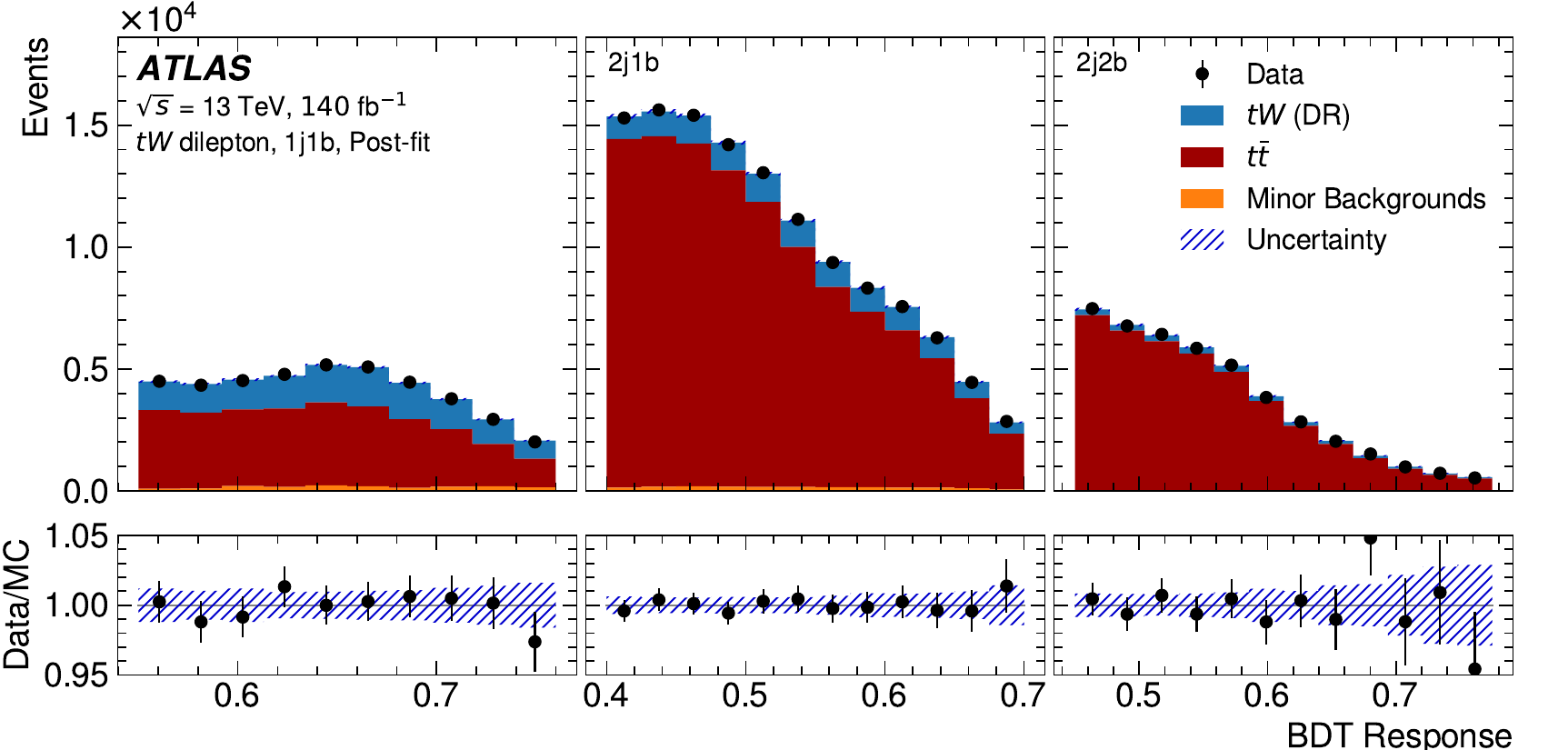}
\caption{The post-fit BDT response distributions for the three regions used in the fit, after applying the additional BDT selection criteria shown in \cref{fig:prefit_dists}.
\capuncertaintybandpostfit
}
\label{fig:postfit_dists}
\end{figure}

The post-fit BDT response distributions are shown in \cref{fig:postfit_dists}.
The post-fit uncertainties are significantly smaller compared with those in the pre-fit
distributions of \cref{fig:prefit_dists} due to the correlations between sources of
systematic uncertainty and constraints imposed by the data.

The impact of the uncertainties in the \tW signal strength, broken down into major categories, is summarized in \cref{tab:syst_table}.
The statistical uncertainty in the measurement is estimated by performing the fit after fixing all nuisance parameters to their post-fit values.
The total systematic uncertainty is calculated by subtracting in quadrature the statistical component of the uncertainty from the total.

\begin{table}[htbp]
\caption{
Impact of relative uncertainties in the \tW cross-section $\sigma_{\tW}$, broken down into major categories.
For each category the impact is calculated by performing a fit where the nuisance parameters in the category are fixed to their best-fit values, and then subtracting the resulting uncertainty in the parameter of interest in quadrature from the uncertainty from the nominal fit.
The symmetric impact is given for simplicity.
\enquote{MC statistical uncertainty} is derived from the nuisance parameters associated with the Poisson distribution terms which represent the effect of the limited size of the MC samples.
\enquote{Data statistical uncertainty} is derived as the subtraction in quadrature of the \enquote{Total uncertainty} and \enquote{Total systematic uncertainty}.
The total systematic uncertainty is not equal to the sum in quadrature of the individual contributions
due to correlations resulting from the fit.
}%
\label{tab:syst_table}
\centering
\sisetup{round-mode=places, round-precision=1, scientific-notation=fixed, fixed-exponent=0}
\begin{tabular}{l
*1{S[table-format=4.1]@{$\,$}S[table-format=2.1]}
}
\toprule{Uncertainty source}                                  & {$\Delta\sigma_{\tW}/\sigma_{\tW}$ [\%]} \\
\midrule
\ttbar modeling                                       & 13.2004    \\ %
Jet energy scale                                      & 11.9887    \\ %
\MET reconstruction and calibration                   & 11.0301    \\ %
\tW modeling                                          & 7.86728    \\ %
Jet energy resolution                                 & 6.99137    \\ %
Jet flavor tagging                                    & 3.74988    \\ %
\Pileup                                               & 2.50143    \\ %
Lepton ($e$ and $\mu$) reconstruction and calibration & 1.85758    \\ %
Other background modeling                             & 0.902007   \\ %
Luminosity                                            & 0.840419   \\ %
PDF (\tW and \ttbar)                                  & 0.588083   \\ %
MC statistical uncertainty                            & 4.65115    \\ %
\midrule
Total systematic uncertainty                          & 19.2419    \\ %
Data statistical uncertainty                          & 1.40616    \\ %
\midrule
Total uncertainty                                     & 19.2932    \\ %
\bottomrule
\end{tabular}
\end{table}

\FloatBarrier
\subsection{Constraints on $|f_\text{LV}V_{tb}|$}

The inclusive cross-section depends on the magnitude of the CKM matrix element $V_{tb}$.
Assuming that \tW-channel production through $|V_{ts}|$ and $|V_{td}|$ is small, the ratio of the measured cross-section to the theoretical prediction is equal to $|f_\text{LV}V_{tb}|^2$, where the left-handed form factor $f_\text{LV}$ could be modified by new physics or radiative corrections through anomalous coupling contributions.
By assuming that the CKM matrix elements $|V_{td}|$ and $|V_{ts}|$ are much smaller than $|V_{tb}|$
and a $V-A$ interaction in the $Wtb$ vertex,
the measured cross-section gives $|f_\text{LV}V_{tb}| = 0.97 \pm 0.10$, consistent with the SM prediction of unity.
In addition to the uncertainty in the measurement, the uncertainty in the \tW cross-section calculation is taken into account and is found to be negligible.


\FloatBarrier
%
%
%

%
\section{Conclusion}%
\label{sec:conclusion}

A measurement of the \tW cross-section using data collected with the ATLAS detector from 2015 to 2018
in $pp$ collisions at a center-of-mass energy of $\sqrt{s} = \SI{13}{\TeV}$ delivered by the LHC
is presented.
The full data sample corresponds to an integrated luminosity of $\SI{140}{\ifb}$.
The \tW final state used for this measurement comprises two oppositely charged leptons ($e^\pm \mu^\mp$ events) and missing transverse momentum from the two $W$-boson decays, and one $b$-jet from the top-quark decay.
BDTs are used to separate the signal from the dominant \ttbar background.

The cross-section for the production of a \Wboson boson in association with a single top quark is measured to be
\begin{equation*}
\sigma_{\tW} = \fittWXSValue\numpmerr{\fittWXSErrorUp}{\fittWXSErrorDn}~\text{pb} = \fittWXSValue
\pm\fittWXSErrorStat\,\text{(stat.)}
\numpmerr{\fittWXSErrorSystUp}{\fittWXSErrorSystDn}\,\text{(syst.)}
\pm\fittWXSErrorLumi\,\text{(lumi.)}\,\unit{\pb},
\end{equation*}
which is in good agreement with the SM prediction of
\(\sigma_{\text{\tW}}^{\text{theory}} = 79.3 \numpmerr{+1.9}{-1.8}\,(\text{scale}) \pm 2.2\,(\text{PDF})\,\unit{\pb}\).
The uncertainty in the measured cross-section is reduced by around \SI{40}{\%} compared with the previous ATLAS measurement using a partial Run~2 dataset.
The stringent constraints imposed on the nuisance parameters of both the \tW DR vs DS uncertainty and the \ttbar parton shower uncertainty are largely reduced by excluding bins of the BDT response in the fit compared with the previous analysis, which leads to a more reliable measurement at the cost of a certain degradation of the precision.
The measured cross-section allows  a direct extraction of the form factor times the CKM matrix element, $|f_\text{LV}V_{tb}| = 0.97 \pm 0.10$, which is consistent with the SM prediction of unity.
%
%


\FloatBarrier

\section*{Acknowledgements}
%

%
%

%
%

We thank CERN for the very successful operation of the LHC and its injectors, as well as the support staff at
CERN and at our institutions worldwide without whom ATLAS could not be operated efficiently.

The crucial computing support from all WLCG partners is acknowledged gratefully, in particular from CERN, the ATLAS Tier-1 facilities at TRIUMF/SFU (Canada), NDGF (Denmark, Norway, Sweden), CC-IN2P3 (France), KIT/GridKA (Germany), INFN-CNAF (Italy), NL-T1 (Netherlands), PIC (Spain), RAL (UK) and BNL (USA), the Tier-2 facilities worldwide and large non-WLCG resource providers. Major contributors of computing resources are listed in Ref.~\cite{ATL-SOFT-PUB-2023-001}.

We gratefully acknowledge the support of ANPCyT, Argentina; YerPhI, Armenia; ARC, Australia; BMWFW and FWF, Austria; ANAS, Azerbaijan; CNPq and FAPESP, Brazil; NSERC, NRC and CFI, Canada; CERN; ANID, Chile; CAS, MOST and NSFC, China; Minciencias, Colombia; MEYS CR, Czech Republic; DNRF and DNSRC, Denmark; IN2P3-CNRS and CEA-DRF/IRFU, France; SRNSFG, Georgia; BMBF, HGF and MPG, Germany; GSRI, Greece; RGC and Hong Kong SAR, China; ISF and Benoziyo Center, Israel; INFN, Italy; MEXT and JSPS, Japan; CNRST, Morocco; NWO, Netherlands; RCN, Norway; MNiSW, Poland; FCT, Portugal; MNE/IFA, Romania; MESTD, Serbia; MSSR, Slovakia; ARIS and MVZI, Slovenia; DSI/NRF, South Africa; MICIU/AEI, Spain; SRC and Wallenberg Foundation, Sweden; SERI, SNSF and Cantons of Bern and Geneva, Switzerland; NSTC, Taipei; TENMAK, T\"urkiye; STFC/UKRI, United Kingdom; DOE and NSF, United States of America.

Individual groups and members have received support from BCKDF, CANARIE, CRC and DRAC, Canada; CERN-CZ, FORTE and PRIMUS, Czech Republic; COST, ERC, ERDF, Horizon 2020, ICSC-NextGenerationEU and Marie Sk{\l}odowska-Curie Actions, European Union; Investissements d'Avenir Labex, Investissements d'Avenir Idex and ANR, France; DFG and AvH Foundation, Germany; Herakleitos, Thales and Aristeia programmes co-financed by EU-ESF and the Greek NSRF, Greece; BSF-NSF and MINERVA, Israel; NCN and NAWA, Poland; La Caixa Banking Foundation, CERCA Programme Generalitat de Catalunya and PROMETEO and GenT Programmes Generalitat Valenciana, Spain; G\"{o}ran Gustafssons Stiftelse, Sweden; The Royal Society and Leverhulme Trust, United Kingdom.

In addition, individual members wish to acknowledge support from Armenia: Yerevan Physics Institute (FAPERJ); CERN: European Organization for Nuclear Research (CERN PJAS); Chile: Agencia Nacional de Investigaci\'on y Desarrollo (FONDECYT 1230812, FONDECYT 1230987, FONDECYT 1240864); China: Chinese Ministry of Science and Technology (MOST-2023YFA1605700), National Natural Science Foundation of China (NSFC - 12175119, NSFC 12275265, NSFC-12075060); Czech Republic: Czech Science Foundation (GACR - 24-11373S), Ministry of Education Youth and Sports (FORTE CZ.02.01.01/00/22\_008/0004632), PRIMUS Research Programme (PRIMUS/21/SCI/017); EU: H2020 European Research Council (ERC - 101002463); European Union: European Research Council (ERC - 948254, ERC 101089007), Horizon 2020 Framework Programme (MUCCA - CHIST-ERA-19-XAI-00), European Union, Future Artificial Intelligence Research (FAIR-NextGenerationEU PE00000013), Italian Center for High Performance Computing, Big Data and Quantum Computing (ICSC, NextGenerationEU); France: Agence Nationale de la Recherche (ANR-20-CE31-0013, ANR-21-CE31-0013, ANR-21-CE31-0022, ANR-22-EDIR-0002), Investissements d'Avenir Labex (ANR-11-LABX-0012); Germany: Baden-Württemberg Stiftung (BW Stiftung-Postdoc Eliteprogramme), Deutsche Forschungsgemeinschaft (DFG - 469666862, DFG - CR 312/5-2); Italy: Istituto Nazionale di Fisica Nucleare (ICSC, NextGenerationEU), Ministero dell'Università e della Ricerca (PRIN - 20223N7F8K - PNRR M4.C2.1.1); Japan: Japan Society for the Promotion of Science (JSPS KAKENHI JP22H01227, JSPS KAKENHI JP22H04944, JSPS KAKENHI JP22KK0227, JSPS KAKENHI JP23KK0245); Netherlands: Netherlands Organisation for Scientific Research (NWO Veni 2020 - VI.Veni.202.179); Norway: Research Council of Norway (RCN-314472); Poland: Ministry of Science and Higher Education (IDUB AGH, POB8, D4 no 9722), Polish National Agency for Academic Exchange (PPN/PPO/2020/1/00002/U/00001), Polish National Science Centre (NCN 2021/42/E/ST2/00350, NCN OPUS nr 2022/47/B/ST2/03059, NCN UMO-2019/34/E/ST2/00393, UMO-2020/37/B/ST2/01043, UMO-2021/40/C/ST2/00187, UMO-2022/47/O/ST2/00148, UMO-2023/49/B/ST2/04085); Slovenia: Slovenian Research Agency (ARIS grant J1-3010); Spain: Generalitat Valenciana (Artemisa, FEDER, IDIFEDER/2018/048), Ministry of Science and Innovation (MCIN \& NextGenEU PCI2022-135018-2, MICIN \& FEDER PID2021-125273NB, RYC2019-028510-I, RYC2020-030254-I, RYC2021-031273-I, RYC2022-038164-I), PROMETEO and GenT Programmes Generalitat Valenciana (CIDEGENT/2019/027); Sweden: Swedish Research Council (Swedish Research Council 2023-04654, VR 2018-00482, VR 2022-03845, VR 2022-04683, VR 2023-03403, VR grant 2021-03651), Knut and Alice Wallenberg Foundation (KAW 2018.0157, KAW 2018.0458, KAW 2019.0447, KAW 2022.0358); Switzerland: Swiss National Science Foundation (SNSF - PCEFP2\_194658); United Kingdom: Leverhulme Trust (Leverhulme Trust RPG-2020-004), Royal Society (NIF-R1-231091); United States of America: U.S. Department of Energy (ECA DE-AC02-76SF00515), Neubauer Family Foundation.

%
%


%
%
\printbibliography
\clearpage
 
\begin{flushleft}
\hypersetup{urlcolor=black}
{\Large The ATLAS Collaboration}

\bigskip

\AtlasOrcid[0000-0002-6665-4934]{G.~Aad}$^\textrm{\scriptsize 104}$,
\AtlasOrcid[0000-0001-7616-1554]{E.~Aakvaag}$^\textrm{\scriptsize 17}$,
\AtlasOrcid[0000-0002-5888-2734]{B.~Abbott}$^\textrm{\scriptsize 123}$,
\AtlasOrcid[0000-0002-0287-5869]{S.~Abdelhameed}$^\textrm{\scriptsize 119a}$,
\AtlasOrcid[0000-0002-1002-1652]{K.~Abeling}$^\textrm{\scriptsize 56}$,
\AtlasOrcid[0000-0001-5763-2760]{N.J.~Abicht}$^\textrm{\scriptsize 50}$,
\AtlasOrcid[0000-0002-8496-9294]{S.H.~Abidi}$^\textrm{\scriptsize 30}$,
\AtlasOrcid[0009-0003-6578-220X]{M.~Aboelela}$^\textrm{\scriptsize 45}$,
\AtlasOrcid[0000-0002-9987-2292]{A.~Aboulhorma}$^\textrm{\scriptsize 36e}$,
\AtlasOrcid[0000-0001-5329-6640]{H.~Abramowicz}$^\textrm{\scriptsize 154}$
% ...
\bigskip
\\

% ...
$^{17}$Department for Physics and Technology, University of Bergen, Bergen; Norway.\\
% ...
$^{30}$Physics Department, Brookhaven National Laboratory, Upton NY; United States of America.\\
% ...
$^{36}$$^{(a)}$Facult\'e des Sciences Ain Chock, Universit\'e Hassan II de Casablanca;$^{(b)}$Facult\'{e} des Sciences, Universit\'{e} Ibn-Tofail, K\'{e}nitra;$^{(c)}$Facult\'e des Sciences Semlalia, Universit\'e Cadi Ayyad, LPHEA-Marrakech;$^{(d)}$LPMR, Facult\'e des Sciences, Universit\'e Mohamed Premier, Oujda;$^{(e)}$Facult\'e des sciences, Universit\'e Mohammed V, Rabat;$^{(f)}$Institute of Applied Physics, Mohammed VI Polytechnic University, Ben Guerir; Morocco.\\
% ...
$^{45}$Physics Department, Southern Methodist University, Dallas TX; United States of America.\\
% ...
$^{50}$Fakult\"{a}t Physik , Technische Universit{\"a}t Dortmund, Dortmund; Germany.\\
% ...
$^{56}$II. Physikalisches Institut, Georg-August-Universit\"{a}t G\"ottingen, G\"ottingen; Germany.\\
% ...
$^{104}$CPPM, Aix-Marseille Universit\'e, CNRS/IN2P3, Marseille; France.\\
% ...
$^{119}$$^{(a)}$New York University Abu Dhabi, Abu Dhabi;$^{(b)}$United Arab Emirates University, Al Ain; United Arab Emirates.\\
% ...
$^{123}$Homer L. Dodge Department of Physics and Astronomy, University of Oklahoma, Norman OK; United States of America.\\
% ...
$^{154}$Raymond and Beverly Sackler School of Physics and Astronomy, Tel Aviv University, Tel Aviv; Israel.\\
% ...
$^{a}$ Also Affiliated with an institute covered by a cooperation agreement with CERN.\\
$^{b}$ Also at An-Najah National University, Nablus; Palestine.\\
$^{c}$ Also at Borough of Manhattan Community College, City University of New York, New York NY; United States of America.\\
$^{d}$ Also at Center for Interdisciplinary Research and Innovation (CIRI-AUTH), Thessaloniki; Greece.\\
$^{e}$ Also at Centro Studi e Ricerche Enrico Fermi; Italy.\\
$^{f}$ Also at CERN, Geneva; Switzerland.\\
% ...

\end{flushleft}


%

%
%

%

%
%
%
%

\end{document}